\documentclass[aps,prd,nofootinbib,twocolumn,superscriptaddress,preprintnumbers,floatfix]{revtex4-2} 

\usepackage[nodisplayskipstretch]{setspace}
\usepackage{cancel}
\usepackage{mathtools}
\usepackage{xspace}
\usepackage{amssymb}
\usepackage{amsmath}
\usepackage{epsfig}
\usepackage{hyperref}
\usepackage{breakurl}
\usepackage{bm}
\usepackage{color}
\usepackage{slashed}
\usepackage{tikz}
\usepackage{xspace}
\usetikzlibrary{decorations.markings}
\usepackage[english]{babel}
\allowdisplaybreaks[4]
\usepackage{needspace}
\usepackage{etoolbox}

\preto{\section}{\par\needspace{\baselineskip}}

\newcommand\beq{\begin{equation}}
\newcommand\eeq{\end{equation}}
\def\bea{\begin{eqnarray}}
\def\eea{\end{eqnarray}}

\DeclareRobustCommand{\SkipTocEntry}[4]{}

\newcommand{\nn}{\nonumber}

\newcommand\beal{\begin{aligned}}
\newcommand\eeal{\end{aligned}}

\newcommand\dd{{\mathrm d}}
\usepackage{xcolor}

\newcommand{\bel}{{\boldsymbol \ell}}

\newcommand{\bn}{{\boldsymbol n}}
\newcommand{\bp}{{\boldsymbol p}}

\newcommand{\bq}{{\boldsymbol q}}

\newcommand{\bI}{{\boldsymbol I}}

\newcommand\cO{\mathcal{O}}

\newcommand{\cG}{\mathcal {G}}

\DeclareRobustCommand{\spiredtt}{%
  \begin{tikzpicture}[
      baseline=(b.base),
      inner sep=0pt,
      outer sep=0pt
    ]
    \node (b) at (0,0) {\textbf{S}};
    \node at (0.2,-0.028) {\textbf{\scalebox{0.75}{P}}};
    \node at (0.35,0) {\textbf{I}};
    \node at (0.52,-0.028) {\textbf{\scalebox{0.75}{\scalebox{-1}[1]{D}}}};
    \node at (0.705,-0.028) {\textbf{\scalebox{0.75}{\scalebox{-0.75}[1]{E}}}};
    \node at (0.92,0) {\textbf{\scalebox{-1}[1]{R}}};
  \end{tikzpicture}\xspace%
}

\makeatother

\begin{document}

\preprint{DESY\, 26-055\\\phantom{~}}
\title{Nonlocal-in-time tail effects in gravitational scattering \\ [0.2cm] to fifth Post-Minkowskian and tenth self-force orders} 
\author{Christoph Dlapa}
\affiliation{ II.~Institute for Theoretical Physics, Hamburg University, D-22761 Hamburg, Germany}

\author{Gregor K\"alin}
\affiliation{ Institute for Theoretical Physics, ETH Z\"urich, 8093 Z\"urich, Switzerland}

\author{Zhengwen Liu}
\affiliation{School of Physics and Shing-Tung Yau Center, Southeast University, Nanjing 210096, China}

\author{Rafael A. Porto}
\affiliation{ Deutsches Elektronen-Synchrotron DESY, Platanenallee 6, 15738 Zeuthen, Germany.}

\begin{abstract}
Using the worldline effective field theory formalism, we derive the nonlocal-in-time conservative contributions arising from tail effects in gravitational scattering to fifth Post-Minkowskian (5PM) and tenth self-force (10SF) orders. The result features multiple polylogarithms of up to weight~three. This challenging computation relies on state-of-the-art integration techniques, including a novel integration-by-parts algorithm: the ``Sparse Integral Reducer'' (\spiredtt). We~find perfect agreement in the overlap with all existing literature through sixth post-Newtonian order. The results presented here provide a key ingredient for isolating the local-in-time component of the conservative two-body dynamics of binary inspirals at 5PM order.
\end{abstract}
\maketitle

\section{Introduction}  The advent of gravitational-wave (GW) astronomy, together with ambitious future observatories such as LISA~\cite{LISA}, Cosmic Explorer~\cite{CE} and the Einstein Telescope~\cite{ET}, has restored the relativistic two-body problem in General Relativity into a central role~\cite{music}. Much of the recent progress is organized around two complementary viewpoints. On the one hand, in the Post-Newtonian (PN) regime, we compute observables simultaneously through an expansion in the relative velocity and in $G$, Newton's constant. On the other hand, in the Post-Minkowskian (PM) approximation, we fix instead an order in $G$ but retain full velocity dependence. In the last several years, both approaches have progressed at an exceptional pace, see e.g.~\cite{Damour:2014jta,tail,Maia:2017yok,Maia:2017gxn,Marchand:2017pir,Foffa:2019rdf,nrgr4pn2,Cho:2022syn,Blanchet:2023sbv,hered1,5pn1,5pn2,hered2,Blumlein:2020pyo,Bini:2020wpo,binidam1,binidam2,Khalil:2022ylj,Blumlein:2021txj,memory,Almeida:2025nic,Cho:2021mqw,Amalberti:2024jaa,Bini:2024rsy,damour1,Damour:2017zjx,bohr,cheung, donal, zvi1,paper1,paper2,b2b3,Damour:2019lcq,pmeft,3pmeft,tidaleft,pmefts,Jakobsen:2021zvh,Mougiakakos:2021ckm,Gabriele2,eftrad,Jakobsen:2022psy,4pmeft,4pmeft2,4pmeftot,4pmzvi,4pmzvi2,Bini:2022enm,Damgaard:2023ttc,dklp,cy,Frellesvig:2023bbf,Bern:2024vqs,Bern:2024adl,Bini:2024rsy,Buonanno:2024vkx,4pmeftloc,5pmeft1loc,Cheung:2024byb,Bini:2024tft,Driesse:2024xad,Driesse:2024feo,Heissenberg:2025ocy,Heissenberg:2025fcr,Bini:2025vuk,Bern:2025wyd,Blanchet:2026suq,Driesse:2026qiz}, with the combination of worldline effective field theory (WEFT) methodologies~\cite{nrgr,nrgrs,dis1,dis2,nrgrss,nrgrs2,iragrg,review,Goldberger:2022ebt} and multiloop integration toolkits developed in collider physics, e.g.~\cite{dklp,cy,Frellesvig:2023bbf,Bern:2024vqs}, playing a major role.\vskip 4pt

From the weak-field and slow-velocity perspective, the dynamics is known in full through 4PN order~\cite{Damour:2014jta,tail,Maia:2017yok,Maia:2017gxn,Marchand:2017pir,Foffa:2019rdf,nrgr4pn2,Cho:2022syn,Blanchet:2023sbv}, while at 5PN both the potential sector and nonlinear radiation-reaction and hereditary contributions are under control~\cite{hered1,5pn1,5pn2,hered2,Blumlein:2021txj,memory,memory2}, together with partial knowledge also at higher orders~\cite{Bini:2020wpo,binidam1,binidam2,Brunello:2025gpf}. In the PM regime, that is particularly natural for scattering computations, the total spacetime impulse has been obtained through 4PM~\cite{4pmeft2,4pmeftot}, including spin effects~\cite{Jakobsen:2023hig,Jakobsen:2023ndj}. At 5PM order, the calculations have been carried out in concert with a self-force (SF) expansion in a small mass ratio, and pushed through 5PM/2SF order~\cite{Driesse:2024xad,Driesse:2024feo,Driesse:2026qiz}. Yet, the 5PM/2SF scattering results obtained in \cite{Driesse:2026qiz} are fixed only up to the value of two boundary integrals in  the `memory' region, reflecting the subtleties associated with nonlinear conservative effects \cite{memory2}.\vskip 4pt

For all of this progress, a conceptual obstruction still remains when attempting to convert scattering information into observables for gravitationally-bound systems. The full PM results of~\cite{4pmeft2,4pmeftot,Driesse:2024xad,Driesse:2024feo,Driesse:2026qiz}, by themselves, do not yield the correct bound dynamics upon analytic continuation via the ``boundary-to-bound" (B2B) dictionary~\cite{paper1,paper2,b2b3}. The source of this failure, however, is well understood: nonlocal-in-time tail terms~\cite{Damour:2014jta,tail} that arise from the scattering of the emitted GW radiation off the binary's own long-range (static) field.\footnote{Other type of nonlocal-in-time hereditary contributions, such as {\it tail-of-tail} effects, also contribute to the conservative part of the dynamics \cite{Bini:2025vuk}. We do not compute these terms here.} These hereditary effects obstruct the naive use of unbound data to describe generic elliptic motion.\vskip 4pt  A resolution of this problem within the PM domain was recently developed in~\cite{4pmeftloc}, and extended to the 5PM/1SF level in~\cite{5pmeft1loc}. Starting from the complete 4PM~\cite{4pmeft2,4pmeftot} and 5PM/1SF~\cite{Driesse:2024xad,Driesse:2024feo} sectors, the universal (local) part of the conservative dynamics was isolated via an explicit calculation of nonlocal tail contributions to gravitational scattering. The purpose of the present paper is to carry this program substantially further. Instead of truncating at the leading SF order, we derive the ${\cal O}(G^5)$ nonlocal-in-time tail corrections to gravitational scattering through tenth order in the mass-ratio expansion.\vskip 4pt

 Following~\cite{dklp,4pmeftloc,5pmeft1loc}, our calculations are organized using the methodology of differential equations combined with a novel integration-by-parts (IBP) algorithm: the ``Sparse Integral Reducer" (\spiredtt), designed to handle the large (absolute) powers of linear propagators involved due to the mass-ratio expansion. The nonlocal tail contributions to the deflection angle to 5PM/10SF can then be shown to depend on multiple polylogarithms (MPLs) up to weight three. We find agreement in the overlap with state-of-the-art derivations in the PN literature through 6PN order~\cite{binidam2}. The results presented here supply one of the key inputs needed to isolate the local-in-time contribution to the conservative dynamics of binary inspirals through 5PM/10SF order. We briefly comment on the necessary steps and will return to this issue elsewhere.\vskip 4pt

\section{Nonlocal tail effects}  

At the technical level, nonlocal tail-type contributions are governed by an integral over the (source) energy spectrum multiplied by a logarithm of the center-of-mass GW frequency~\cite{b2b3,binidam2}. The nonlocal part of the radial action then takes the form
\beq
 {\cal S}^{(\rm nloc)}_{r\,} = -\frac{G E}{2\pi}   \int_{-\infty}^{+\infty} \frac{d\omega}{2\pi} \frac{dE_{\rm src}}{d\omega}  \log \left[\left(2e^{\gamma_E}GM\omega\right)^2\right]\,.\label{nloc1}
\eeq   
We adopt the conventions of~\cite{4pmeftloc}: $E$ denotes the total energy, $\frac{dE_{\rm src}}{d\omega}$ is the (odd-in-velocity) source GW spectrum in the center-of-mass frame, $M=m_1+m_2$ is the total mass, and $\gamma_E$ is Euler's constant. The scattering angle satisfies $\frac{\chi}{2\pi}=-\partial_J{\cal S}_r$, with $J$ the total angular momentum. Writing the conservative deflection angle in the PM approximation, we parameterize its coefficients as
\beq
\frac{\chi}{2} = \sum^{n=5}_{n=1} \left(\chi_b^{(n)} + \chi_b^{(n)\log} \log \frac{\hat b}{\Gamma}\right) \hat b^{-n},\label{eq:angle}
\eeq 
to leading order in the tail expansion. We introduced $\hat b\equiv b/GM$, with $b$ the impact parameter, $\gamma\equiv u_1\!\cdot u_2$, with $u_a$ ($a=1,2$) the incoming four-velocities, and use a mostly negative metric convention. We also defined $\Gamma\equiv E/M=\sqrt{1+2\nu(\gamma-1)}$, with $\nu\equiv m_1m_2/M^2$.\vskip 4pt

 While the total expression for ratio $\chi/\Gamma$ is known to truncate in the $\nu$ expansion~\cite{Damour:2019lcq,paper1}, that is not the case for the nonlocal part. This is due to integrand in \eqref{nloc1} carrying nontrivial dependence on both the relative velocity and the mass ratio~\cite{4pmeftloc}. Therefore, we adopt the same strategy as in~\cite{5pmeft1loc}, and expand the integrand in powers of the small mass ratio, $m_2/m_1\ll 1$, which is directly associated with the traditional SF expansion. As already shown in~\cite{4pmeftloc}, the logarithmic contribution admits an SF-exact representation, which was already obtained at 5PM order in~\cite{5pmeft1loc}.We compute here the remaining nonlocal tail contributions to $\chi_b^{(5)}/\Gamma$ through 10SF orders.

\section{Building the Integrand}

The integrand for the nonlocal part of the radial action~\eqref{nloc1} has already been constructed in~\cite{5pmeft1loc}.
Using a partial fraction decomposition, together with relabeling symmetries, we embed all the relevant scalar integrals into a single integral family. The family is defined by three delta functions in $\{\ell_1\cdot u_1,\, \ell_2\cdot u_1,\, \ell_3\cdot u_2\}$, a set of linear propagators, $\{\ell_1\cdot u_2,\ell_2\cdot u_2,\ell_3\cdot u_1\}$,
and completed by a choice of square propagators, $\big\{-\ell_1^2,-\ell_2^2,-\ell_3^2,-(\ell_1-q)^2,-(\ell_2-q)^2,-(\ell_3-q)^2, -(\ell_1-\ell_2)^2,-(\ell_2-\ell_3)^2,-(\ell_3-\ell_1)^2\big\}\,.$
The nontrivial mass dependence enters through the factor of $\log\left[\omega^2\right]=\log\left[(k\cdot u_\textrm{com})^2\right]$, with \beq u_{\rm com} \equiv \frac{m_1 u_1+m_2 u_2}{M\Gamma},\eeq which multiplies the square and linear propagators. We~identify two regions of integration each involving a {\it single} radiation mode characterized by the momentum $k$ going on-shell (which we denote as the `1rad' region~\cite{dklp}), with $k_{(1)}=\ell_2-\ell_3$ and $k_{(2)}=\ell_1-\ell_3$. As it turns out, the presence of two independent regions plays an important role in the computation of the relevant integrals. \vskip 4pt 
Upon expansion in the small mass ratio $m_2/m_1\ll 1$,
\begin{equation}
\label{eq:log-SF-expansion}
  \begin{aligned}
   & \log\left[(k\cdot u_\textrm{com})^2\right] = \log\left[(k\cdot u_1)^2\right]\\
    &\quad - 2\sum_{n=1}^\infty \frac{(-1)^n}{n} \left(\frac{k\cdot u_2}{k\cdot u_1}\right)^n \left(\frac{m_2}{m_1}\right)^n \\ &\qquad -2\log\left[\left(1+\frac{m_2}{m_1}\right)\Gamma\right]\,,
  \end{aligned}
\end{equation}
such that the computation can then be split into three separate parts according to the three lines of~\eqref{eq:log-SF-expansion}. The logarithm in the last line is independent of the loop momenta and therefore simply multiplies the total radiated energy computed in~\cite{4pmeftot}. The contribution of the logarithm in the first line was obtained in~\cite{5pmeft1loc} by treating the latter as a standard linear propagator with non-integer power, i.e.~$(k\cdot u_1)^{2\tilde{\epsilon}}$, and successively keeping only the part linear in $\tilde{\epsilon}$. (It is convenient to distinguish $\tilde\epsilon$ from the more standard $\epsilon \equiv (3-d)/2$ introduced in dimensional regularization.)  The main goal of this paper is to compute the contributions from the additional terms in the second line up to an overall 10SF order.  
  
\section{Sparse Integral Reducer}

After the full answer is written as a series of Feynman integrals, we exploit linear relations to write them in terms of a set of master integrals. This not only vastly reduces the total number of integrals that needs to be computed, but also facilitates the subsequent computation of the master integrals through the method of differential equations. This so-called IBP reduction is by now a standard procedure in high-energy physics and implemented in various publicly available computer codes, such as \texttt{Kira}~\cite{Lange:2025fba}, \texttt{Fire}\cite{Smirnov:2025prc}, and \texttt{LiteRed}~\cite{Lee:2013mka}.\vskip 4pt 
At leading order in the SF expansion \cite{5pmeft1loc}, the \texttt{Kira} algorithm was able to deal with the powers of $\tilde{\epsilon}$ in the exponents after some minor modifications. Yet, although the scalar integrals belong to the same family as those considered in~\cite{4pmeftot,dklp} for the computation of the impulse, the higher order corrections in the mass ratio involve increasingly large powers of the linear propagators, which reach $\pm 12$ at 10SF order. In this regime, a naive Laporta-style reduction becomes cumbersome, as the large index ranges translate directly into substantial runtime and memory demands. A different strategy was therefore needed to be able to progress beyond the leading SF order. \vskip 4pt 

In order to address these issues is that we introduced the new \spiredtt program. The latter is built on a modified version of the strategy implemented in \texttt{LiteRed}~\cite{Lee:2013mka}. Unlike the traditional Laporta approach, \spiredtt constructs \emph{symbolic} reduction rules at the level of sectors, rather than solving only for a predetermined list of target integrals. The derivation of these rules is largely independent of the specific subset of integrals that ultimately appears in the calculation: any integral within the covered sectors can be reduced subsequently by straightforward rule application. This feature is particularly valuable at higher orders in the perturbative expansion, where the full set of required integrals is not known \emph{a priori} in closed form and increases significantly with each~order.\vskip 4pt

The advantage of symbolic rules is that they effectively remove the ``forward solving'' (and seeding) stage, which often constitutes a major bottleneck in Laporta implementations. Once symbolic reduction rules are available, a minimal forward-solved (upper triangular) form of the IBP system can be constructed by recursively traversing the dependency tree associated with the set of target integrals. In our setup, master integrals are identified precisely as those to which no symbolic rule applies, \textit{i.e.} the integrals corresponding to columns without a pivot element. The precise form and size of the forward-solved system depend strongly on the chosen integral ordering.\footnote{By incorporating ideas  to improve the symbolic reduction rules similar to those in~\cite{delaCruz:2026mas}---which can be implemented straightforwardly within the \spiredtt algorithm---we can in principle reduce the size of the forward-solved system even further.}\vskip 4pt 

The computational challenge is shifted to the \emph{application} of the symbolic rules and the reduction of large integral expressions. For this purpose, we developed a dedicated engine based on a bottom-up strategy for sparse back substitution, using finite-field arithmetic (FF) and rational reconstruction (RR). More concretely, reductions are performed over suitable prime fields, where algebraic operations are fast and the growth of intermediate expressions is tightly controlled. The resulting data are then combined via RR to recover the exact rational coefficients. This approach allows us to navigate an intermediate reduction graph containing order $10^8$ integrals, while explicitly computing the associated reduction coefficients. In practice, FF\&RR techniques ensure that the computational cost is dominated by modular numerical back substitution of a sparse matrix and by RR, rather than by large rational symbolic arithmetic at intermediate stages. This made it possible to perform the computation entirely in main memory, without resorting to disk-based storage. Moreover, the numerical reduction can be straightforwardly parallelized, with negligible overhead and minimal need for synchronization.\vskip 4pt 

Finally, at tenth order in the expansion, we also encounter additional sectors beyond those appearing at leading order in~\cite{5pmeft1loc}. Although conceptually straightforward, this introduces a number of new masters.  This enlargement turns out to be important, as it requires extending the rule-generation step to cover the relevant topologies. Once these are incorporated, the full set can be treated uniformly within the same framework. The \spiredtt algorithm successfully generated symbolic rules for all sectors needed in our calculation.

\section{Integration}

For the computation of the master integrals we follow the approach outlined in detail in \cite{dklp}: We derive differential equations in $x$, given by $\gamma=\frac{1}{2}\left(x+\frac{1}{x}\right)$ and transform them to canonical form \cite{Henn:2013pwa}. The differential equations posses the same structure as the ones we encountered before in~\cite{4pmeftot,dklp}, with the exception of a few extra master integrals. We note that, for the leading term in the SF expansion, it is advantageous to temporarily set $\tilde{\epsilon}=\eta\epsilon$ before deriving canonical differential equations that are factorized in $\epsilon$. For this task we use private codes which deal with the $\eta$ parameter. \vskip 4pt 
 The solution of the differential equations depend on MPLs up to transcendental weight three and two, for the leading and higher-order SF  terms, respectively.\footnote{The increased weight of the leading term may be attributed to the fact that the logarithm in the first line of~\eqref{eq:log-SF-expansion} has weight one, thus increasing the maximum weight relative to the other terms. Furthermore, elliptic integrals are completely absent, since they appear only in the direction of the impact parameter at ${\cal O}(G^4)$.} The MPLs are defined through the relation
\beq
\begin{aligned}
G(a_1,\ldots, a_n;z) &= \int_0^{z} \frac{dt}{t-a_1} G(a_2,\ldots,a_n;t)\,,\\\label{eq:G}
 G(\cdot ;z)&=1\,, \quad G(\underbrace{0,\ldots, 0}_{n};z) = \tfrac{1}{n!} \log^n z\,,
 \end{aligned}
 \eeq
 with $(a_n,z)$ complex numbers. The result can be written in terms of a five letter {\it alphabet} $a_i\in\{0,1,2, 1\pm i\}$. We~choose the variable $z=1-x$, which has a direct connection with the PN expansion, and furthermore it allows us to efficiently determine an independent set of boundary constants using the methods developed in~\cite{dklp}.\vskip 4pt 

The integration problem then reduces to the computation of boundary conditions in the $z\rightarrow 0$ $(\gamma \to 1)$ limit, characterized in our case by a single propagator going on-shell. An expansion in small velocity leads to a factorization of the full three-loop integrals into an (outer) two-loop integral times an (inner) tadpole, which only talk to each other through the dependence on the argument of a single linear propagator $\bel_i\cdot \bn$. The $\bel_i$ is one of the two outer integration momenta, and $\bn$ is chosen such that $\bel\cdot \bn = \ell_z$.\vskip 4pt The tadpole integral is known analytically for any power of the single propagator, and evaluates to a result proportional to $(\bel_i\cdot \bn)^d$. The dependence on $d$, as well as the introduction of the $\tilde\epsilon$ parameter to handle the leading logarithmic-dependent term, implies non-integer powers for one of the linear propagators of the outer integrals, which can all be embedded into the following family
\begin{equation}
  \begin{aligned}
  &\hspace{3cm} \bI^{\pm\pm}_{a_1a_2;a_3 a_4 a_5 a_6 a_7}=\\
  &\int \frac{e^{2 \gamma_\textrm{E}\epsilon}}{\pi^d}\frac{\dd^d \bel_1 \dd^d\bel_2}{(\bel_1\cdot \bn\pm i0)^{a_1}(\bel_2\cdot \bn \pm i0)^{a_2}(\bel_1^2)^{a_3}(\bel_2^2)^{a_4}}\\
  &\qquad\times\frac{1}{((\bel_1-\bq)^2)^{a_5}((\bel_2-\bq)^2)^{a_6}((\bel_1-\bel_2)^2)^{a_7}}\,,
  \end{aligned}
\end{equation}
with $\bq\cdot \bn=0$, $a_{i\neq 1}$ integer numbers, and $a_1 = \mathbb{Z} + {\cal O}(\epsilon,\tilde \epsilon)$.\vskip 4pt 
This set of integrals fulfills similar IBP identities as their $(d+1)$-dimensional counterparts. The algorithm implemented in \spiredtt is able to produce a complete set of symbolic rules that allows a reduction of all relevant integrals in this family into the set of masters: 
  \begin{align}
    &  \bI^{-+}_{(1+2\epsilon-2\tilde\epsilon)1;10011}\,, \,\, \bI^{+\pm}_{(2\epsilon-2\tilde\epsilon)0;10011}\,,  \,\, \bI^{++}_{(1+2\epsilon-2\tilde\epsilon)1;10011}\,,\nn \\
    &\bI^{-+}_{(2\epsilon)1;10011}\,, \,\, \bI^{-+}_{(1+2\epsilon)1;10011}\,, \,\,\bI^{-+}_{(1+2\epsilon)1;10101}\,,\nn\\
    &\bI^{+\pm}_{(2\epsilon)0;10011}\,, \,\,\bI^{++}_{(2\epsilon)1;10011}\,, \,\,\bI^{+\pm}_{(-1+2\epsilon)0;10011}\,,\nn\\
    &\bI^{++}_{(1+2\epsilon)1;10011}\,, \,\,\bI^{++}_{(1+2\epsilon)1;10101}\,,
  \end{align}
which are sufficient to all SF orders. We~find complete analytic results through a Schwinger parametrization and iterated integration of the Schwinger parameters.
Some examples of similar boundary integrals---including a thorough discussion of many of the finer details---can be found~in~\cite{dklp,Jinno:2022sbr}. For the readers' convenience we list their values, and $\epsilon$-expanded forms, in Appendix~\ref{app1}.

\section{Scattering Data}

Combining the leading (non-logarithmic) SF contribution from~\cite{5pmeft1loc} with the newly obtained results in this paper, we find the following structure
\beq\begin{aligned}\label{chinlocnlog}
\frac{\chi_{b(\rm nloc)}^{(5)(10\rm SF)}}{\Gamma \nu}  &= 
  \sum_{i=1}^{26} \left[h_i^{(0)}(x) + h_i^{(1)}(x) \nu^1\right.\\
   & \left.+ \sqrt{1-4\nu}h_i^{\left(\small{1/2}\right)}(x) + \Delta h_i(x,\nu)\right] \cG_i(x)\,,
    \end{aligned}
\eeq
for the coefficient of the scattering angle in~\ref{eq:angle}, where the $h_i^{\left(0,1,\small{1/2}\right)}$ are rational functions of $x$ only, and $\Delta h_i(x,\nu)$ are rational in $x$ and polynomial in $\nu$, starting at ${\cal O}(\nu^2)$ and up to ${\cal O}(\nu^9)$. Furthermore, the $\Delta h_i(x,\nu)$ vanish except for $i=1,3,4,6,7,8$. See the ancillary file for (computer-readable) explicit expressions. The combinations of MPLs in the~$\cG_i(x)$'s are listed in Appendix~\ref{app2}. \vskip 4pt 

The result in \eqref{chinlocnlog}, together with the SF-exact logarithmic pieces \cite{5pmeft1loc}, completes the knowledge of nonlocal tail-like contributions to 5PM/10SF order.  For the sake of comparison, expressing the scattering angle including the logarithmic terms as a function of the reduced angular momentum, $\chi= \sum_n 2\chi_j^{(n)}j^{-n}$, with $j=J/(G M^2 \nu)$, $v_{\infty} = \sqrt{\gamma^2-1}$, and expanding to 10PN order we find 
\begin{widetext}
  {
    \small
    \begin{align} \label{10pn}
      \chi_{j\textrm{(nloc)}}^{(5)\rm (10PN)}&= \frac{\nu v_\infty^3}{2} \Bigg\{
      -\frac{5440 }{27}  -\frac{25088}{45}  \log \left(\frac{2 v_\infty}{\sqrt{j}}\right) + v_\infty^2 \left[\left(\frac{55808 \nu}{45}-\frac{297728}{525}\right) \log \left(\frac{2 v_\infty}{\sqrt{j}}\right)+\frac{155200 \nu}{189}+\frac{374368}{2625}\right]\nn\\
      &\quad+ v_\infty^4 \left[\left(-\frac{86528 \nu^2}{45}+\frac{1152896 \nu}{1575}-\frac{3525568}{11025}\right) \log \left(\frac{2 v_\infty}{\sqrt{j}}\right)-\frac{7750528 \nu^2}{4725}-\frac{41361872 \nu}{165375}+\frac{166150088}{385875}\right]\nn\\
      &\quad+ v_\infty^6 \left[\left(\frac{117248 \nu^3}{45}-\frac{875008 \nu^2}{1575}+\frac{3096256 \nu}{6615}+\frac{2429664}{13475}\right) \log \left(\frac{2 v_\infty}{\sqrt{j}}\right)\right.\nn\\
        &\qquad\qquad\left.+\frac{1046272 \nu^3}{405}+\frac{21978496 \nu^2}{33075}-\frac{22499132024 \nu}{38201625}+\frac{1137056876}{7640325}\right]\nn\\
      &\quad+ v_\infty^8 \left[\left(-\frac{147968 \nu^4}{45}+\frac{3968 \nu^3}{105}-\frac{886048 \nu^2}{1323}-\frac{279650936 \nu}{363825}+\frac{1620350092}{2477475}\right) \log \left(\frac{2 v_\infty}{\sqrt{j}}\right)\right.\nn\\
        &\qquad\qquad\left.-\frac{80567008 \nu^4}{22275}-\frac{2743988288 \nu^3}{1819125}+\frac{204678995432 \nu^2}{420217875}-\frac{3137954156509 \nu}{5462832375}+\frac{38172272288071}{120182312250}\right]\\
      &\quad+ v_\infty^{10} \left[\left(\frac{178688 \nu^5}{45}+\frac{1293568 \nu^4}{1575}+\frac{1289632 \nu^3}{1225}+\frac{16506176 \nu^2}{11025}-\frac{307880492 \nu}{315315}-\frac{735017339222}{676350675}\right) \log \left(\frac{2 v_\infty}{\sqrt{j}}\right)\right.\nn\\
        &\qquad\qquad+\frac{1367324512 \nu^5}{289575}+\frac{18434632624 \nu^4}{6449625}+\frac{75486806288 \nu^3}{2630252625}+\frac{88618604658512 \nu^2}{71016820875}-\frac{308380410342341 \nu}{426100925250}\nn\\
        &\qquad\qquad\left.-\frac{81515761234704829}{121864864621500}\right]\nn\\
      &\quad+ v_\infty^{12} \left[\left(-\frac{209408 \nu^6}{45}-\frac{3184256 \nu^5}{1575}-\frac{57721472 \nu^4}{33075}-\frac{98860192 \nu^3}{40425}+\frac{48577741364 \nu^2}{52026975}+\frac{525179938397 \nu}{225450225}\right.\right.\nn\\
        &\qquad\qquad\left.+\frac{108812096826773}{68987768850}\right) \log \left(\frac{2 v_\infty}{\sqrt{j}}\right)
        -\frac{1704352288 \nu^6}{289575}-\frac{338314320616 \nu^5}{70945875}-\frac{245498965780408 \nu^4}{213050462625}\nn\\
        &\qquad\qquad\left.-\frac{486269791848056 \nu^3}{213050462625}+\frac{2009480696060767 \nu^2}{2343555088875}+\frac{25168487123370619421 \nu}{12430216191393000}+\frac{27550186358971497707}{24860432382786000}\right]\Bigg\}\,.\nn
    \end{align}
}\end{widetext}

As stressed in \cite{4pmeftloc}, the nonlocal-in-time contribution defined in \cite{binidam1,binidam2} contains, besides \eqref{nloc1}---denoted $W_1$ in \cite{binidam2}---an additional term ($W_2$) given by a time integral involving $\frac{dE}{dt}$ weighted by $\log r(t)$. This (coordinate dependent) term arises from quasi-instantaneous, potential-region interactions, and therefore we do not assign it to the nonlocal-in-time sector. Upon comparing the expression in \eqref{10pn} with the deflection angle obtained from the $W_1$-only result of \cite{binidam2}, we find perfect agreement in the overlapping regime of validity to 6PN. The PN-expanded value for the nonlocal-in-time ($W_1$-only) contribution to the scattering angle at ${\cal O}(G^5)$ through 30PN/10SF order is provided in the ancillary file.

\section{Towards the 5PM gravitational dynamics} 

The identification of nonlocal-in-time effects in the scattering angle allows us to in principle isolate the universal part of the conservative dynamics entirely within the PM framework. This strategy was followed in our previous work at 4PM \cite{4pmeftloc} and 5PM/1SF \cite{5pmeft1loc} order, respectively. Using the B2B dictionary \cite{paper1,paper2,b2b3}, the resulting Hamiltonian applicable for elliptic motion  through 5PM order can be separated into local- and nonlocal-in-time contributions as follows \cite{5pmeft1loc}
\beq\begin{aligned}
\label{Htot5pm} 
\hat H^{\rm ell}_{\rm 5PM}
&= \hat E_0
+\sum_{i=1}^{i=5} \frac{\hat c_{i\rm (loc)}}{\hat r^i}
+\sum_{i=1}^{i=5} \frac{\hat c_{i\rm (nloc)}}{\hat r^i}  \\
&- \sum_{i=4}^{i=5}
\left. G \left(\hat H\frac{dE_{\rm src}}{dt}\right)\right|_{(i-1)\rm PM}
\log \left(\frac{\hat r}{e^{2\gamma_E}}\right) \,, 
\end{aligned}
\eeq
with $\hat H \equiv \tfrac{H}{M\nu}, \hat r = \tfrac{r}{GM}$, and $\hat E_0 \equiv  \tfrac{1}{M\nu}\sum_a \sqrt{\bp^2+m^2_a}$ is the normalized kinetic term. The $\hat c_{1|2|3(\rm loc)}$ are the PM-exact coefficients through ${\cal O}(G^3)$ computed in \cite{cheung,pmeft,zvi1,3pmeft}, with  $\gamma = (E_1E_2+\bp^2)/m_1m_2$.\vskip 4pt  The coefficient $\hat c_{4(\rm loc)}$ was
obtained in \cite{4pmeftloc} to all SF orders, while the 5PM coefficient $\hat c_{5(\rm loc)}$ was derived in \cite{5pmeft1loc} at the 1SF level. The factor multiplying $\log \left(\tfrac{\hat r}{e^{2\gamma_E}}\right)$ in \eqref{Htot5pm} can be shown to depend on lower order $\hat c_i$'s combined with the source (``1rad"-only) energy flux. The relation follows from the general structure in \eqref{nloc1}, including the factor of $GE$, together
with the cancellation in dimensional regularization of ultraviolet and
infrared divergences---and associated $\log b$ terms---between local-
and nonlocal-in-time contributions \cite{tail,b2b3,4pmeft,4pmeftloc}. For the
bound Hamiltonian to 5PM order this gives the structure:
\bea
\hat c_{5(\rm tot)}^{\rm ell, \log} \label{Hlog}
&=& -G\left. \left( \hat H \frac{dE_{\rm src}}{dt}\right)\right|_{G^4}\\\,
&=& -G \left(
\left. \frac{\hat c_1}{\hat r}\frac{dE_{\rm src}}{dt}\right|_{G^3}
+ \left.\hat E_0 \frac{dE_{\rm src}}{dt}\right|_{G^4}
\right) \,,\nn
\eea
which is displayed in \eqref{Htot5pm}. The explict values for the source energy fluxes are given in \cite{4pmeft,4pmeftloc} at ${\cal O}(G^3)$, and the 4PM component can be found in \cite{5pmeft1loc}.\vskip 4pt  The remaining non-logarithmic contributions,
$\hat c_{i\rm (nloc)}$, are not known in a PM framework for elliptic motion. Closely related coefficients, albeit in a different gauge, have been computed in the PN approximation via a small-eccentricity
expansion through 6PN order \cite{binidam1,binidam2}. These results can be easily translated to an isotropic gauge by extending the analysis performed in \cite{5pmeft1loc}.\vskip 4pt

Armed with the nonlocal-in-time correction to the scattering angle at 5PM/10SF order computed in this paper, and following the same sequence of manipulations as in~\cite{4pmeftloc,5pmeft1loc}, it would in principle be straightforward to extract the corresponding contribution to $\hat c_{5(\rm loc)}$ through the same 5PM/10SF order, provided the value of the \emph{full} conservative scattering angle at ${\cal O}(G^5)$ is at hand. Significant progress in this direction was recently reported in~\cite{Driesse:2026qiz}. However, while the potential and tail-like regions are under control, several subtleties associated with memory-like contributions prevent the conservative value of the total scattering angle from being fully settled \cite{memory2}.\vskip 4pt Nevertheless, it is still useful to split the 5PM coefficient into two pieces, 
\beq
  \hat c_{5(\rm loc)}
  =
  \hat c^{\rm (P+T)}_{5(\rm loc)}
  +
  \hat c^{\rm (M)}_{5(\rm loc)}\,,
\eeq
where the potential-plus-tail part, $\hat c^{\rm (P+T)}_{5(\rm loc)}$, starts at ${\cal O}(\nu)$, and the memory-like piece, $\hat c^{\rm (M)}_{5(\rm loc)}$, enters at ${\cal O}(\nu^2)$. Using the results in~\cite{Driesse:2026qiz}, we can then extend our analysis in~\cite{5pmeft1loc} and determine $\hat c^{\rm (P+T)}_{5(\rm loc)}$ through 10SF order. The remaining memory-like contribution can be incorporated through a PN-expanded representation \cite{memory2},  in parallel with the nonlocal tail sector.\vskip 4pt 

Although the Hamiltonian in \eqref{Htot5pm} is engineered for the bound problem, there is in principle a caveat to this program. Namely, the local-in-time universal component must be analytically continued from the regime of hyperbolic encounters, where the potential-region contribution to the scattering calculation at ${\cal O}(G^5\nu^2)$ contains spurious poles at $\gamma=3$~\cite{Driesse:2026qiz}. If taken at face value, these singularities would feed directly into $\hat c^{\rm (P+T)}_{5(\rm loc)}$ starting at 2SF order, and cannot be removed by supplementing it with a PN-truncated expression for $\hat c^{\rm (M)}_{5(\rm loc)}$.\vskip 4pt  The authors of~\cite{Driesse:2026qiz} therefore advocated for an alternative definition of the conservative dynamics, the so-called ``$\gamma\text{-}3$'' prescription, in which a memory-like contribution is constructed as to cancel the singular behavior arising from the potential region. However, as discussed in~\cite{memory2}, this prescription introduces an unnatural sign flip between tail and memory contributions associated with the same term in the effective action and moreover, if implemented at higher PM orders, it leads to spurious $\pi^2$ terms at ${\cal O}(G^6\nu^2)$. Alternatively, inspired by Feynman's $i0^+$ prescription, a different routing was advocated in \cite{memory2} for the memory-like region at 5PM order. Yet, although devoid of $\epsilon$ poles, Feynman's prescription does not appear to remove the singular behavior at $\gamma=3$. \vskip 4pt

We envision two possible routes for resolving these issues. Firstly, as advocated in~\cite{memory2}, one natural option is to extend the isotropic-like representation to the full 5PM dynamics. This would require going beyond the impulse alone and deriving, within the same framework, the associated energy and angular-momentum losses as well as the time delay. Such a construction would be independent of any prior split into conservative and dissipative sectors, and should therefore be free of spurious singularities. In this setting, the nonlocal-in-time tail effects constructed here could also be systematically removed and reinserted at the level of the full dynamics within a PN-expanded framework.\vskip 4pt 
A second possibility is to retain the split into conservative and dissipative contributions, while defining the conservative sector strictly within the  domain relevant for bound motion. Since the latter is restricted to $\gamma<1$, the singular behavior at $\gamma=3$ is never  attained. From this perspective, an all-orders-in-velocity expression for $\hat c^{\rm (P+T)}_{5(\rm loc)}$ may still provide an improved description of the two-body bound dynamics---after all, potential interactions find their natural habitat in such regime. Yet, it should not be interpreted as an object that can be naively extrapolated to the relativistic limit of unbound motion. We will return to these considerations elsewhere.
\section{Conclusions}  

The combination of modern multiloop integration technology with the WEFT formalism has opened a new analytic window onto gravitational scattering in General Relativity, making it possible to compute (gauge-invariant) observables with increasing precision~\cite{4pmeft2,4pmeftot,Driesse:2024feo,Driesse:2026qiz}. Yet the passage from hyperbolic encounters to bound inspirals is not automatic. As demonstrated in~\cite{paper1,paper2,b2b3}, the instantaneous pieces and logarithmic sector possess the expected universal character, whereas the complete scattering answer contains genuinely orbit-dependent nonlocal-in-time information. Without disentangling this hereditary component, the hyperbolic result cannot be directly used to model binaries in nearly circular or mildly eccentric configurations.\vskip 4pt  

The first systematic implementation of such a disentangling within the PM expansion was achieved in~\cite{4pmeftloc} at ${\cal O}(G^4)$, where the local dynamics was separated from tail-induced nonlocal effects to all orders in the SF expansion. Building on that construction, in \cite{5pmeft1loc} we extended the decomposition of hereditary terms in the conservative scattering angle at 5PM/1SF order. Subtracting this piece from the total value in~\cite{Driesse:2024xad} yields a genuinely instantaneous part of the 5PM dynamics at 1SF order which can then be mapped through the B2B dictionary to generic bound motion \cite{paper1,paper2}.\vskip 4pt

The purpose of the present paper was to carry this program further by deriving the ${\cal O}(G^5)$ nonlocal-in-time tail corrections to gravitational scattering through 10SF order. This provides a significantly more detailed characterization of the hereditary sector at 5PM, and supplies the information needed to in principle disentangle, order by order in $\nu$, the local-in-time dynamics from the genuinely nonlocal contribution to the scattering angle. Explicit formulae for the relevant building blocks, together with PN expansions through 30th order, are provided in the ancillary material.\vskip 4pt 

The results in this paper also pave the path forward to construct the 5PM two-body Hamiltonian for bound motion through 10SF order. The main remaining issue lies in the treatment of memory-like effects, and in particular how the $\gamma=3$ singularity induced by potential interactions is ultimately handled.  A natural resolution, as advocated in \cite{memory2}, may require either formulating the full dynamics without splitting into conservative and dissipative pieces, or else interpreting the (resummed) conservative sector strictly within the PN domain of elliptic orbits, where singularities at $\gamma >1$ remain outside the physical regime of bound motion. We will address these issues in  more detail in future work.\vskip 4pt 

Apart from their immediate application to waveform modelling---that is particular relevant for third-generation GW detectors \cite{CE,ET,LISA}---the plethora of recent results pushing the boundaries of our understanding of gravitational interactions underscores the maturity of the WEFT framework~\cite{iragrg,review,Goldberger:2022ebt} as a systematic approach to the two-body problem in general relativity. In~combination with differential-equation methods, IBP reduction, and related integration tools~\cite{dklp,cy,Frellesvig:2023bbf}, it offers a systematic route to navigate through the increasingly intricate integration problem that governs the relativistic binary dynamics. In this direction, the newly developed \spiredtt program provides a powerful computational engine for future calculations, and we expect it to play a central role in unlocking further breakthroughs at higher PM orders.
\newpage
{\bf Acknowledgements}. G.K. and R.A.P. would like to thank the participants of the Nordita program ``Amplitudes, Strong-Field Gravity and Resummation" for discussions. The work of C.D. is funded by the Deutsche Forschungsgemeinschaft (DFG, German Research Foundation) - Project number 555751285. R.A.P. acknowledges support by the DFG under Germany's Excellence Strategy -- EXC 2121 ``Quantum Universe" -- 390833306. The work of Z.L. was supported partially by the European Union's Horizon Europe research and innovation program under the Marie Sk\l{}odowska-Curie grant agreement No 101146918 `GWtheory' and by the Start-up Research Fund of Southeast University No.\,RF1028624160. 
\appendix

\begin{widetext}

\section{Boundary integrals}\label{app1}

We list here the results for the required boundary integrals. The answers contain generalized hypergeometric functions of type $\,_4F_3(z)$ with $\epsilon$-dependent arguments evaluated at $z=1$:
\begingroup
\allowdisplaybreaks
    {
      \small
      \begin{align}
        \bI^{-+}_{(2\epsilon)1;10011} &= \frac{i \pi ^2 4^{4 \epsilon+1} e^{2 \gamma_\textrm{E}  \epsilon+i \pi  \epsilon} \sin (\pi  \epsilon) (\csc (2 \pi  \epsilon)+\csc (4 \pi  \epsilon)) \Gamma \left(\frac{1}{2}-3 \epsilon\right) \Gamma (6 \epsilon)  }{\Gamma \left(\frac{3}{2}-2 \epsilon\right) \Gamma (-\epsilon) \Gamma (4 \epsilon+1)} \, _4F_3\left(
        \begin{matrix}
          1-4 \epsilon,\frac{1}{2}-3 \epsilon,\frac{1}{2}-2 \epsilon,\frac{1}{2}-\epsilon\\
          1-6 \epsilon,1-2 \epsilon,\frac{3}{2}-2 \epsilon
        \end{matrix}
        \,\middle|\,1\right)\nn\\
        &\quad -\frac{\pi ^{7/2} \left(4^{6 \epsilon}-1\right) e^{2 \gamma_\textrm{E}  \epsilon} \epsilon (\tan (\pi  \epsilon)-i) \csc (4 \pi  \epsilon) (\sec (\pi  \epsilon)+\sec (3 \pi  \epsilon))}{\Gamma (1-4 \epsilon) \Gamma \left(\frac{1}{2}-2 \epsilon\right) \Gamma (2 \epsilon+1)^2 \sinh(\epsilon \log (64))}\, _4F_3\left(
        \begin{matrix}
          \frac{1}{2},1-2 \epsilon,\frac{1}{2}-\epsilon,\epsilon+\frac{1}{2}\\
          \frac{3}{2},1-4 \epsilon,2 \epsilon+1
        \end{matrix}
        \,\middle|\,1\right)\nn\\
        &\quad-\frac{i \pi  64^{\epsilon} e^{2 \gamma_\textrm{E}  \epsilon+i \pi  \epsilon} \cos (4 \pi  \epsilon) \Gamma \left(\frac{1}{2}-2 \epsilon\right) \Gamma \left(4 \epsilon+\frac{1}{2}\right)}{\epsilon-2 \epsilon \cos (2 \pi  \epsilon)}\,,\nn\\
        \bI^{-+}_{(1+2\epsilon)1;10011} &= \frac{i \pi ^{3/2} 16^{\epsilon} e^{2 (\gamma_\textrm{E} +i \pi ) \epsilon} \Gamma (-3 \epsilon) \Gamma \left(2 \epsilon+\frac{1}{2}\right) \Gamma (6 \epsilon+1) (2 \cos (2 \pi  \epsilon )-1) (1+i \cot (\pi  \epsilon )) }{\Gamma (1-\epsilon) \Gamma (2 \epsilon+1)^2}\, _4F_3\left(
        \begin{matrix}
          \frac{1}{2}-\epsilon,-4 \epsilon,-3 \epsilon,-2 \epsilon\\
          \frac{1}{2}-2 \epsilon,1-2 \epsilon,-6 \epsilon
        \end{matrix}
        \,\middle|\,1\right)\nn\\
        &\quad+\frac{4^{3 \epsilon+1} e^{2 \gamma_\textrm{E}  \epsilon+i \pi  \epsilon} \sin (8 \pi  \epsilon) \cos (\pi  \epsilon)  \Gamma \left(-2 \epsilon-\frac{1}{2}\right) \Gamma \left(\frac{1}{2}-\epsilon\right) \Gamma (\epsilon+1) \Gamma \left(4 \epsilon+\frac{1}{2}\right) } {\sqrt{\pi } (2 \cos (2 \pi  \epsilon)+1)\sin (4 \pi  \epsilon)} \, _4F_3\left(
        \begin{matrix}
          \frac{1}{2},\frac{1}{2}-2 \epsilon,\frac{1}{2}-\epsilon,\epsilon+1\\
          \frac{3}{2},\frac{1}{2}-4 \epsilon,2 \epsilon+\frac{3}{2}
        \end{matrix}
        \,\middle|\,1\right)\nn\\
        &\quad+\frac{\pi ^{3/2} 2^{4 \epsilon+1} e^{2 \gamma_\textrm{E}  \epsilon+i \pi  \epsilon} \Gamma (-\epsilon) \left(2 \cos (\pi  \epsilon) \Gamma (-2 \epsilon) \Gamma (4 \epsilon+1)+\sqrt{\pi } 16^{\epsilon} \csc (3 \pi  \epsilon) \Gamma \left(2\epsilon+\frac{1}{2}\right)\right)}{\Gamma \left(\frac{1}{2}-2 \epsilon\right) \Gamma (\epsilon+1)}\,,\nn\\
        \bI^{-+}_{(1+2\epsilon)1;10101} &= 0\,,\nn\\
        \bI^{+\pm}_{(2\epsilon)0;10011} &= \frac{\pi ^2 4^{\epsilon} e^{2 \gamma_\textrm{E}  \epsilon-i \pi  \epsilon} \csc (3 \pi  \epsilon) \Gamma \left(\frac{1}{2}-2 \epsilon\right) \Gamma\left(\frac{1}{2}-\epsilon\right)}{\Gamma \left(\frac{3}{2}-5 \epsilon\right) \Gamma (1-\epsilon) \Gamma \left(\epsilon+\frac{1}{2}\right)}\,,\nn\\
        \bI^{++}_{(2\epsilon)1;10011} &= -\frac{i \pi ^2 2^{8 \epsilon+1} e^{2 \gamma_\textrm{E}  \epsilon-i \pi  \epsilon} (2 \cos (2 \pi  \epsilon)+1) \Gamma \left(\frac{1}{2}-3 \epsilon\right) \Gamma (6 \epsilon)}{(\cos (\pi \epsilon)+\cos (3 \pi  \epsilon)) \Gamma \left(\frac{3}{2}-2 \epsilon\right) \Gamma (-\epsilon) \Gamma (4 \epsilon+1)}\, _4F_3\left(
        \begin{matrix}
          1-4 \epsilon,\frac{1}{2}-3 \epsilon,\frac{1}{2}-2 \epsilon,\frac{1}{2}-\epsilon\\
          1-6 \epsilon,1-2 \epsilon,\frac{3}{2}-2 \epsilon
        \end{matrix}
        \,\middle|\,1\right)\nn\\
        &\quad+\frac{i \pi ^{7/2} 2^{6 \epsilon+1} e^{2\gamma_\textrm{E}  \epsilon-3 i \pi  \epsilon} \left(1+e^{4 i \pi  \epsilon}\right) }{(\sin (2 \pi  \epsilon)+\sin (6 \pi  \epsilon)+\sin (8 \pi \epsilon)) \Gamma \left(\frac{1}{2}-2 \epsilon\right) \Gamma (-4 \epsilon) \Gamma (2 \epsilon+1)^2}
        \, _4F_3\left(
        \begin{matrix}
          \frac{1}{2},1-2 \epsilon,\frac{1}{2}-\epsilon,\epsilon+\frac{1}{2}\\
          \frac{3}{2},1-4 \epsilon,2 \epsilon+1
        \end{matrix}
        \,\middle|\,1\right)\nn\\
        &\quad-\frac{i \pi  64^{\epsilon} e^{2 \gamma_\textrm{E}  \epsilon-i \pi \epsilon} \cos (4 \pi  \epsilon) \Gamma \left(\frac{1}{2}-2 \epsilon\right) \Gamma \left(4 \epsilon+\frac{1}{2}\right)}{\epsilon-2 \epsilon \cos (2 \pi  \epsilon)}\,,\nn\\
        \bI^{+\pm}_{(-1+2\epsilon)0;10011} &= -\frac{i \pi  4^{\epsilon} e^{2 \gamma_\textrm{E}  \epsilon-i \pi  \epsilon} \sin (\pi  \epsilon) \sec (3 \pi  \epsilon) \Gamma (1-2 \epsilon) \Gamma \left(\frac{1}{2}-\epsilon\right)}{\Gamma \left(\frac{5}{2}-5 \epsilon\right)}\,,\\
        \bI^{++}_{(1+2\epsilon)1;10011} &= -\frac{3\ 2^{-4 \epsilon} e^{2 \gamma_\textrm{E}  \epsilon-i \pi  \epsilon} \sin (2 \pi  \epsilon) \Gamma \left(\frac{1}{2}-2 \epsilon\right) \Gamma (-3 \epsilon) \Gamma (\epsilon) \Gamma (6 \epsilon) }{\sqrt{\pi }} \, _4F_3\left(
        \begin{matrix}
          \frac{1}{2}-2 \epsilon,-4 \epsilon,-2 \epsilon,2 \epsilon+1\\
          \frac{1}{2}-3 \epsilon,1-2 \epsilon,1-\epsilon
        \end{matrix}
        \,\middle|\,1\right)\nn\\
        &\quad+\frac{\pi  4^{4 \epsilon+1} e^{2 \gamma_\textrm{E}  \epsilon-i \pi  \epsilon} \sin (2 \pi  \epsilon) \csc (3 \pi  \epsilon) \Gamma (-2 \epsilon) \Gamma \left(2 \epsilon+\frac{1}{2}\right)}{\epsilon \Gamma \left(\frac{1}{2}-2 \epsilon\right)} \, _4F_3\left(
        \begin{matrix}
          \frac{1}{2}-\epsilon,-3 \epsilon,-\epsilon,3 \epsilon+1\\
          \frac{1}{2}-2 \epsilon,1-\epsilon,\epsilon+1
        \end{matrix}
        \,\middle|\,1\right)\nn\\
        &\quad-\frac{\pi ^{3/2} 8^{2 \epsilon+1} e^{2 \gamma_\textrm{E}  \epsilon-i \pi \epsilon} \Gamma (\epsilon+1) \Gamma \left(5 \epsilon+\frac{3}{2}\right) (\sin (4 \pi  \epsilon) \csc (6 \pi  \epsilon)-1)}{(4 \epsilon+1) \Gamma \left(\epsilon+\frac{3}{2}\right) \Gamma \left(3 \epsilon+\frac{3}{2}\right)}
        \, _4F_3\left(
        \begin{matrix}
          \frac{1}{2}-\epsilon,\epsilon+\frac{1}{2},\epsilon+1,5 \epsilon+\frac{3}{2}\\
          \epsilon+\frac{3}{2},2 \epsilon+\frac{3}{2},3 \epsilon+\frac{3}{2}
        \end{matrix}
        \,\middle|\,1\right)\,,\nn\\
        \bI^{++}_{(1+2\epsilon)1;10101} &= \frac{i \pi ^2 4^{2 \epsilon+1} e^{2 (\gamma_\textrm{E} +i \pi ) \epsilon} \Gamma \left(\frac{1}{2}-\epsilon\right)}{\left(-1+e^{6 i \pi  \epsilon}\right) \epsilon \Gamma \left(\frac{1}{2}-3 \epsilon\right)}\,,\nn\\
        \bI^{-+}_{(1+2\epsilon-2\tilde\epsilon)1;10011} &= \frac{\pi ^{3/2} 4^{1-\epsilon } e^{2 \gamma  \epsilon +i \pi  (\epsilon -\tilde\epsilon)} \Gamma \left(\frac{1}{2}-\epsilon \right) \cos (\pi  (3 \epsilon -\tilde\epsilon))  \Gamma (\tilde\epsilon-3 \epsilon ) \Gamma (6 \epsilon-2 \tilde\epsilon +1)}{\sin (2 \pi  \tilde\epsilon-4 \pi  \epsilon ) \Gamma (2 \tilde\epsilon-4 \epsilon +1) \Gamma (-2 \tilde\epsilon+2 \epsilon +1) \Gamma (-\tilde\epsilon+2 \epsilon +1)} \,_4F_3\left(
        \begin{matrix}
          2 \tilde\epsilon-4 \epsilon ,\tilde\epsilon-3 \epsilon ,\tilde\epsilon-2 \epsilon ,\frac{1}{2}-\epsilon \\
          2 \tilde\epsilon-6 \epsilon ,\tilde\epsilon-2 \epsilon +\frac{1}{2},\tilde\epsilon-2 \epsilon +1
        \end{matrix}\,\middle|\,1\right)\nn\\
        &\quad-\frac{\pi ^2 2^{4 \epsilon-2 \tilde\epsilon +2} e^{2 \gamma  \epsilon +i \pi  (\epsilon -\tilde\epsilon)}  \cos (\pi  (4 \epsilon -\tilde\epsilon)) \Gamma \left(\tilde\epsilon-2 \epsilon -\frac{1}{2}\right) \Gamma \left(4 \epsilon-\tilde\epsilon +\frac{1}{2}\right) }{\sin (\pi (3 \epsilon -\tilde\epsilon)) \cos (\pi  \epsilon )\Gamma \left(\epsilon +\frac{1}{2}\right) \Gamma (\tilde\epsilon-2 \epsilon ) \Gamma \left(-\tilde\epsilon+\epsilon +\frac{1}{2}\right)}\, _4F_3\left(
        \begin{matrix}
          \frac{1}{2},\tilde\epsilon-2 \epsilon +\frac{1}{2},\frac{1}{2}-\epsilon ,\epsilon-\tilde\epsilon +1\\
          \frac{3}{2},\tilde\epsilon-4 \epsilon +\frac{1}{2},-\tilde\epsilon+2 \epsilon +\frac{3}{2}\\
        \end{matrix}\,\middle|\,1\right)\nn\\
        &\quad+\frac{\pi ^{7/2} 2^{-2 \tilde\epsilon+4 \epsilon +1} e^{2 \gamma  \epsilon +i \pi  (\epsilon -\tilde\epsilon)} \csc (\pi  \epsilon ) \sin (\pi  (4 \epsilon -\tilde\epsilon)) \csc (\pi (2 \epsilon -\tilde\epsilon)) \csc (\pi  (3 \epsilon -\tilde\epsilon)) \Gamma (-\tilde\epsilon+4 \epsilon +1)}{\Gamma (\epsilon +1) \Gamma \left(\tilde\epsilon-2 \epsilon +\frac{1}{2}\right) \Gamma (-\tilde\epsilon+\epsilon +1) \Gamma (-\tilde\epsilon+2 \epsilon +1)}
        \,,\nn\\
        \bI^{+\pm}_{(2\epsilon-2\tilde\epsilon)0;10011} &= \frac{\pi ^2 4^{\epsilon} e^{2 \gamma  \epsilon-i \pi  (\epsilon-\tilde\epsilon)} \Gamma \left(\frac{1}{2}-\epsilon\right) \csc (3 \pi  \epsilon-\pi  \tilde\epsilon) \Gamma \left(-2 \epsilon+\tilde\epsilon+\frac{1}{2}\right)}{\Gamma (1-\epsilon) \Gamma \left(\epsilon-\tilde\epsilon+\frac{1}{2}\right) \Gamma \left(-5 \epsilon+2 \tilde\epsilon+\frac{3}{2}\right)}\,,\nn\\
        \bI^{++}_{(1+2\epsilon-2\tilde\epsilon)1;10011} &= -\frac{i \pi ^{5/2} e^{2 \gamma  \epsilon } 2^{-4 \tilde\epsilon+6 \epsilon +2} \csc (\pi  \tilde\epsilon-3 \pi  \epsilon ) \Gamma (2 \tilde\epsilon-4 \epsilon ) \Gamma \left(3 \epsilon-2 \tilde\epsilon +\frac{1}{2}\right) }{\left(e^{2 i \pi  (\epsilon -\tilde\epsilon)} - 1\right) \Gamma (-2 \epsilon ) \Gamma (\tilde\epsilon-2 \epsilon +1) \Gamma (\tilde\epsilon-\epsilon +1) \Gamma (2 \epsilon-2 \tilde\epsilon +1)}\, _4F_3\left(
        \begin{matrix}
          2 \tilde\epsilon-4 \epsilon ,\tilde\epsilon-2 \epsilon ,\tilde\epsilon-2 \epsilon +\frac{1}{2},2 \epsilon +1\\
          2 \tilde\epsilon-3 \epsilon +\frac{1}{2},\tilde\epsilon-2 \epsilon +1,\tilde\epsilon-\epsilon +1
        \end{matrix}\,\middle|\,1\right)\nn\\
        &\quad +\frac{i \pi ^{5/2} e^{2 \gamma  \epsilon } 2^{-2 \tilde\epsilon+4 \epsilon +2} \Gamma (-2 \epsilon ) \csc (\pi  \tilde\epsilon-3 \pi  \epsilon ) \Gamma \left(-\tilde\epsilon+2 \epsilon +\frac{1}{2}\right) }{\left(e^{2 i \pi  (\epsilon -\tilde\epsilon)} -1\right) \Gamma (1-\epsilon ) \Gamma (2 \tilde\epsilon-4 \epsilon ) \Gamma (\epsilon-\tilde\epsilon +1) \Gamma (2 \epsilon-2 \tilde\epsilon +1)}\, _4F_3\left(\begin{matrix}
          \tilde\epsilon-3 \epsilon ,\frac{1}{2}-\epsilon ,-\epsilon ,3 \epsilon-\tilde\epsilon +1\\
          \tilde\epsilon-2 \epsilon +\frac{1}{2},1-\epsilon ,\epsilon-\tilde\epsilon +1
        \end{matrix}\,\middle|\,1\right)\nn\\
        &\quad -\frac{2 \pi ^{5/2} e^{2 \gamma  \epsilon -i \pi  (\epsilon -\tilde\epsilon)} \Gamma \left(\frac{1}{2}-\epsilon \right) \cos (2 \pi  \tilde\epsilon-5 \pi  \epsilon ) \sec (\pi  \tilde\epsilon-2 \pi  \epsilon ) \Gamma \left(-2 \tilde\epsilon+5 \epsilon +\frac{3}{2}\right) }{(\sin (\pi  \tilde\epsilon)-\sin (3 \pi  \tilde\epsilon-6 \pi  \epsilon )) \Gamma (2 \tilde\epsilon-4 \epsilon ) \Gamma \left(\epsilon-\tilde\epsilon +\frac{3}{2}\right) \Gamma \left(2 \epsilon-\tilde\epsilon +\frac{3}{2}\right) \Gamma \left(3 \epsilon -2 \tilde\epsilon+\frac{3}{2}\right)}\nn\\
        &\qquad\times\, _4F_3\left(\begin{matrix}
          \frac{1}{2}-\epsilon ,\epsilon-\tilde\epsilon +\frac{1}{2},\epsilon-\tilde\epsilon +1,5 \epsilon-2\tilde\epsilon +\frac{3}{2}\\
          \epsilon-\tilde\epsilon +\frac{3}{2},-\tilde\epsilon+2 \epsilon +\frac{3}{2},3 \epsilon-2 \tilde\epsilon +\frac{3}{2}
        \end{matrix}\,\middle|\,1\right)
        \,.\nn
      \end{align}
    }
The expansion of these functions for small $(\epsilon, \tilde \epsilon)$ is rather non trivial.
To deal with the two-parameter problem, as explained in the main text we set $\tilde\epsilon = \eta\epsilon$ and expand the boundary conditions for small $\epsilon$. This allows for a systematic cancellation of spurious poles in the integrand.  The resulting values are listed below, with some of them obtained through a high-precision numerical evaluation followed by a reconstruction using the PSLQ algorithm~\cite{bailey1991polynomial,Bailey:1999nv}:
\begin{align}
  \bI^{-+}_{(2\epsilon)1;10011} &=
  \frac{i \pi ^2}{\epsilon}
  -\pi ^2 (\pi -2 i \log (2))
  -\frac{1}{2} i\pi ^2 (\pi -2 i \log (2))^2 \epsilon\nn\\
  &\quad+\frac{1}{6} \pi ^2 \left(8 i \left(\log ^3(2)-140 \zeta (3)\right)+\pi ^3-12 \pi  \log ^2(2)-6 i \pi ^2 \log (2)\right) \epsilon^2
  +\cO(\epsilon^3)
  \,,\nn\\
  \bI^{-+}_{(1+2\epsilon)1;10011} &=
  \frac{5 \pi }{3 \epsilon^2}
  +\frac{5 i \pi ^2}{3 \epsilon}
  -\frac{19 \pi ^3}{9}
  +\cO(\epsilon)
  \,,\nn\\
  \bI^{++}_{(2\epsilon)1;10011} &=
  \frac{i \pi ^2}{\epsilon}
  +\pi ^2 (\pi +2 i \log (2))
  +\left(\frac{3 i \pi ^4}{2}+2 i \pi ^2 \log ^2(2)+\pi ^3 \log (4)\right)\epsilon\\
  &\quad+\frac{1}{6} \pi ^2 \left(8 i \left(\log ^3(2)-56 \zeta (3)\right)+11 \pi ^3+12 \pi  \log ^2(2)+18 i \pi ^2 \log (2)\right)\epsilon^2
  +\cO(\epsilon^3)
  \,,\nn\\
  \bI^{++}_{(1+2\epsilon)1;10011} &=
  -\frac{\pi }{\epsilon^2}
  +\frac{i \pi ^2}{\epsilon}
  +\frac{7 \pi ^3}{9}
  +\cO(\epsilon)
  \,,\nn\\
  \bI^{-+}_{(1+2\epsilon-2\eta\epsilon)1;10011} &=
  -\frac{2 \pi  (\eta -5)}{(\eta -3) (\eta -2) \epsilon ^2}
  +\frac{2 i \pi ^2 (\eta -1) (\eta -5)}{(\eta -3) (\eta -2) \epsilon }
  +\frac{\pi ^3 \left(4 \eta ^3-32 \eta ^2+61 \eta -38\right)}{3 (\eta -3) (\eta -2)}
  +\cO(\epsilon)
  \,,\nn\\
  \bI^{++}_{(1+2\epsilon-2\eta\epsilon)1;10011} &=
  \frac{2 \pi }{(\eta -2) \epsilon ^2}
  +\frac{2 i \pi ^2 (\eta -1)}{(\eta -2) \epsilon }
  +\frac{\pi ^3 \left(4 \eta ^3-32 \eta ^2+61 \eta -38\right)}{3 (\eta -3) (\eta -2)}
  +\cO(\epsilon)
  \,.\nn
\end{align}
\endgroup
After the poles are removed, we can safely reinstate $\eta = \tilde \epsilon / \epsilon$, with final expressions becoming rational functions in both $\epsilon$ and $\tilde\epsilon$ that can be easily expanded near zero.

\section{Basis of special functions}\label{app2}
We list below the set of special functions introduced in the main text depending on MPLs up to weight three:
\begingroup
\allowdisplaybreaks
\begin{align}
  \cG_1(x) &= 1\,,\nn\\
  \cG_2(x) &= G(0;1-x)+G(2;1-x)\,,\nn\\
  \cG_3(x) &= G(1;1-x)\,,\nn\\
  \cG_4(x) &= G(0,1;1-x)\,,\nn\\
  \cG_5(x) &= G(0;1-x) G(1;1-x)-G(0,1;1-x)+G(1,2;1-x)\,,\nn\\
  \cG_6(x) &= \frac{1}{2} G(1;1-x)^2\,,\nn\\
  \cG_7(x) &= G(1-i,1;1-x)+G(1+i,1;1-x)\,,\nn\\
  \cG_8(x) &= G(1;1-x) G(2;1-x)-G(1,2;1-x)\,,\nn\\
  \cG_9(x) &= G(0,0,1;1-x)+\frac{1}{2} G(0,1,1;1-x)\,,\nn\\
  \cG_{10}(x) &= -\frac{1}{2} G(2;1-x) G(1;1-x)^2+G(1,2;1-x) G(1;1-x)+G(0;1-x) G(0,1;1-x)-2 G(0,0,1;1-x)\nn\\
  &\quad-2 G(0,1,1;1-x)+G(0,1,2;1-x)-G(1,1,2;1-x)\,,\nn\\
  \cG_{11}(x) &= -G(0,1,1;1-x)+G(0,1-i,1;1-x)+G(0,1+i,1;1-x)\,,\nn\\
  \cG_{12}(x) &= \frac{1}{2} G(2;1-x) G(1;1-x)^2-G(1,2;1-x) G(1;1-x)-\frac{1}{2} G(0,1,1;1-x)+G(0,2,1;1-x)\nn\\
  &\quad+G(1,1,2;1-x)\,,\nn\\
  \cG_{13}(x) &= G(1;1-x) G(0,1;1-x)\,,\nn\\
  \cG_{14}(x) &= \frac{1}{6} G(1;1-x)^3\,,\nn\\
  \cG_{15}(x) &= \frac{1}{2} G(0;1-x) G(1;1-x)^2-G(2;1-x) G(1;1-x)^2-(G(0,1;1-x)-2 G(1,2;1-x)) G(1;1-x)\nn\\
  &\quad-G(0,1,1;1-x)-G(1,1,2;1-x)\,,\nn\\
  \cG_{16}(x) &= G(1,1-i,1;1-x)+G(1,1+i,1;1-x)\,,\nn\\
  \cG_{17}(x) &= G(1;1-x) (G(1;1-x) G(2;1-x)-G(1,2;1-x))\,,\\
  \cG_{18}(x) &= \frac{1}{2} \left[G(2;1-x) G(1;1-x)^2+(-2 G(1,2;1-x)+G(1-i,1;1-x)+G(1+i,1;1-x)) G(1;1-x)\right.\nn\\
    &\quad\quad\,\,\left.+2 G(0,1,1;1-x)+2 G(1,1,2;1-x)-G(1,1-i,1;1-x)-G(1,1+i,1;1-x)\right]\,,\nn\\
  \cG_{19}(x) &= -\frac{1}{2} G(2;1-x) G(1;1-x)^2+G(1,2;1-x) G(1;1-x)+G(0;1-x)  (G(1-i,1;1-x)+G(1+i,1;1-x))\nn\\
  &\quad-G(0,1,1;1-x)-G(0,1-i,1;1-x)-G(0,1+i,1;1-x)-G(1,1,2;1-x)-G(1-i,0,1;1-x)\nn\\
  &\quad+G(1-i,1,2;1-x)-G(1+i,0,1;1-x)+G(1+i,1,2;1-x)\,,\nn\\
 \cG_{20}(x) &= \frac{1}{2} G(2;1-x) G(1;1-x)^2-G(1,2;1-x) G(1;1-x)+G(0,1,1;1-x)+G(1,1,2;1-x)\nn\\
  &\quad+G(1-i,1-i,1;1-x)+G(1-i,1+i,1;1-x)+G(1+i,1-i,1;1-x)+G(1+i,1+i,1;1-x)\,,\nn\\
  \cG_{21}(x) &= \frac{3}{2} G(2;1-x) G(1;1-x)^2-3 G(1,2;1-x) G(1;1-x)+3 G(0,1,1;1-x)+3 G(1,1,2;1-x)\nn\\
  &\quad+G(1-i,0,1;1-x)+G(1-i,2,1;1-x)+G(1+i,0,1;1-x)+G(1+i,2,1;1-x)\,,\nn\\
  \cG_{22}(x) &= G(2;1-x) G(0,1;1-x)+G(0,1,1;1-x)-G(0,1,2;1-x)-G(0,2,1;1-x)\nn\\
  &\quad+\frac{1}{4} \left[-G(2;1-x) G(1;1-x)^2+2 G(1,2;1-x) G(1;1-x)-2 G(1,1,2;1-x)\right]\,,\nn\\
  \cG_{23}(x) &= -G(2;1-x) G(1;1-x)^2+2 G(1,2;1-x) G(1;1-x)-G(2;1-x) G(0,1;1-x)\nn\\
  &\quad+G(0;1-x) (G(1;1-x) G(2;1-x)-G(1,2;1-x))+G(2;1-x) G(1,2;1-x)-G(0,1,1;1-x)\nn\\
  &\quad+G(0,1,2;1-x)-2 G(1,1,2;1-x)-2 G(1,2,2;1-x)\,,\nn\\
  \cG_{24}(x) &= -\frac{1}{2} G(2;1-x) G(1;1-x)^2+G(1,2;1-x) G(1;1-x)\nn\\
  &\quad+G(2;1-x) (G(1-i,1;1-x)+G(1+i,1;1-x))-G(1,1,2;1-x)-G(1-i,1,2;1-x)\nn\\
  &\quad-G(1-i,2,1;1-x)-G(1+i,1,2;1-x)-G(1+i,2,1;1-x)\,,\nn\\
  \cG_{25}(x) &= \frac{1}{4} \left[G(2;1-x) G(1;1-x)^2+2 \left(G(2;1-x)^2-G(1,2;1-x)\right)  G(1;1-x)\right.\nn\\
    &\quad\quad\,\,\left.-4 G(2;1-x) G(1,2;1-x)+2 G(1,1,2;1-x)+4 G(1,2,2;1-x)\right]\,,\nn\\
  \cG_{26}(x) &= \frac{1}{2} G(1;1-x) G(2;1-x)^2-G(1,2;1-x) G(2;1-x)+G(0,0,1;1-x)+G(1,2,2;1-x)\nn
\end{align}
\endgroup

\end{widetext}

\bibliographystyle{apsrev4-2}
\bibliography{ref4PM}

\begin{thebibliography}{93}%
\makeatletter
\providecommand \@ifxundefined [1]{%
 \@ifx{#1\undefined}
}%
\providecommand \@ifnum [1]{%
 \ifnum #1\expandafter \@firstoftwo
 \else \expandafter \@secondoftwo
 \fi
}%
\providecommand \@ifx [1]{%
 \ifx #1\expandafter \@firstoftwo
 \else \expandafter \@secondoftwo
 \fi
}%
\providecommand \natexlab [1]{#1}%
\providecommand \enquote  [1]{``#1''}%
\providecommand \bibnamefont  [1]{#1}%
\providecommand \bibfnamefont [1]{#1}%
\providecommand \citenamefont [1]{#1}%
\providecommand \href@noop [0]{\@secondoftwo}%
\providecommand \href [0]{\begingroup \@sanitize@url \@href}%
\providecommand \@href[1]{\@@startlink{#1}\@@href}%
\providecommand \@@href[1]{\endgroup#1\@@endlink}%
\providecommand \@sanitize@url [0]{\catcode `\\12\catcode `\$12\catcode
  `\&12\catcode `\#12\catcode `\^12\catcode `\_12\catcode `\%12\relax}%
\providecommand \@@startlink[1]{}%
\providecommand \@@endlink[0]{}%
\providecommand \url  [0]{\begingroup\@sanitize@url \@url }%
\providecommand \@url [1]{\endgroup\@href {#1}{\urlprefix }}%
\providecommand \urlprefix  [0]{URL }%
\providecommand \Eprint [0]{\href }%
\providecommand \doibase [0]{https://doi.org/}%
\providecommand \selectlanguage [0]{\@gobble}%
\providecommand \bibinfo  [0]{\@secondoftwo}%
\providecommand \bibfield  [0]{\@secondoftwo}%
\providecommand \translation [1]{[#1]}%
\providecommand \BibitemOpen [0]{}%
\providecommand \bibitemStop [0]{}%
\providecommand \bibitemNoStop [0]{.\EOS\space}%
\providecommand \EOS [0]{\spacefactor3000\relax}%
\providecommand \BibitemShut  [1]{\csname bibitem#1\endcsname}%
\let\auto@bib@innerbib\@empty
\bibitem [{\citenamefont {Amaro-Seoane}\ \emph {et~al.}(2017)\citenamefont
  {Amaro-Seoane} \emph {et~al.}}]{LISA}%
  \BibitemOpen
  \bibfield  {author} {\bibinfo {author} {\bibfnamefont {P.}~\bibnamefont
  {Amaro-Seoane}} \emph {et~al.} (\bibinfo {collaboration} {LISA}),\
  }\href@noop {} {\  (\bibinfo {year} {2017})},\ \Eprint
  {https://arxiv.org/abs/1702.00786} {arXiv:1702.00786 [astro-ph.IM]}
  \BibitemShut {NoStop}%
\bibitem [{\citenamefont {Reitze}\ \emph {et~al.}(2019)\citenamefont {Reitze}
  \emph {et~al.}}]{CE}%
  \BibitemOpen
  \bibfield  {author} {\bibinfo {author} {\bibfnamefont {D.}~\bibnamefont
  {Reitze}} \emph {et~al.},\ }\href@noop {} {\bibfield  {journal} {\bibinfo
  {journal} {Bull. Am. Astron. Soc.}\ }\textbf {\bibinfo {volume} {51}},\
  \bibinfo {pages} {035} (\bibinfo {year} {2019})},\ \Eprint
  {https://arxiv.org/abs/1907.04833} {arXiv:1907.04833 [astro-ph.IM]}
  \BibitemShut {NoStop}%
\bibitem [{\citenamefont {Abac}\ \emph {et~al.}(2025)\citenamefont {Abac} \emph
  {et~al.}}]{ET}%
  \BibitemOpen
  \bibfield  {author} {\bibinfo {author} {\bibfnamefont {A.}~\bibnamefont
  {Abac}} \emph {et~al.},\ }\href@noop {} {\  (\bibinfo {year} {2025})},\
  \Eprint {https://arxiv.org/abs/2503.12263} {arXiv:2503.12263 [gr-qc]}
  \BibitemShut {NoStop}%
\bibitem [{\citenamefont {Porto}(2017)}]{music}%
  \BibitemOpen
  \bibfield  {author} {\bibinfo {author} {\bibfnamefont {R.~A.}\ \bibnamefont
  {Porto}},\ }\href@noop {} {\  (\bibinfo {year} {2017})},\ \Eprint
  {https://arxiv.org/abs/1703.06440} {arXiv:1703.06440 [physics.pop-ph]}
  \BibitemShut {NoStop}%
\bibitem [{\citenamefont {Damour}\ \emph {et~al.}(2014)\citenamefont {Damour},
  \citenamefont {Jaranowski},\ and\ \citenamefont
  {Sch{\"a}fer}}]{Damour:2014jta}%
  \BibitemOpen
  \bibfield  {author} {\bibinfo {author} {\bibfnamefont {T.}~\bibnamefont
  {Damour}}, \bibinfo {author} {\bibfnamefont {P.}~\bibnamefont {Jaranowski}},\
  and\ \bibinfo {author} {\bibfnamefont {G.}~\bibnamefont {Sch{\"a}fer}},\
  }\href {https://doi.org/10.1103/PhysRevD.89.064058} {\bibfield  {journal}
  {\bibinfo  {journal} {Phys. Rev. D}\ }\textbf {\bibinfo {volume} {89}},\
  \bibinfo {pages} {064058} (\bibinfo {year} {2014})},\ \Eprint
  {https://arxiv.org/abs/1401.4548} {arXiv:1401.4548} \BibitemShut {NoStop}%
\bibitem [{\citenamefont {Galley}\ \emph {et~al.}(2016)\citenamefont {Galley},
  \citenamefont {Leibovich}, \citenamefont {Porto},\ and\ \citenamefont
  {Ross}}]{tail}%
  \BibitemOpen
  \bibfield  {author} {\bibinfo {author} {\bibfnamefont {C.}~\bibnamefont
  {Galley}}, \bibinfo {author} {\bibfnamefont {A.}~\bibnamefont {Leibovich}},
  \bibinfo {author} {\bibfnamefont {R.~A.}\ \bibnamefont {Porto}},\ and\
  \bibinfo {author} {\bibfnamefont {A.}~\bibnamefont {Ross}},\ }\href
  {https://doi.org/10.1103/PhysRevD.93.124010} {\bibfield  {journal} {\bibinfo
  {journal} {Phys. Rev. D}\ }\textbf {\bibinfo {volume} {93}},\ \bibinfo
  {pages} {124010} (\bibinfo {year} {2016})},\ \Eprint
  {https://arxiv.org/abs/1511.07379} {arXiv:1511.07379} \BibitemShut {NoStop}%
\bibitem [{\citenamefont {Maia}\ \emph
  {et~al.}(2017{\natexlab{a}})\citenamefont {Maia}, \citenamefont {Galley},
  \citenamefont {Leibovich},\ and\ \citenamefont {Porto}}]{Maia:2017yok}%
  \BibitemOpen
  \bibfield  {author} {\bibinfo {author} {\bibfnamefont {N.~T.}\ \bibnamefont
  {Maia}}, \bibinfo {author} {\bibfnamefont {C.~R.}\ \bibnamefont {Galley}},
  \bibinfo {author} {\bibfnamefont {A.~K.}\ \bibnamefont {Leibovich}},\ and\
  \bibinfo {author} {\bibfnamefont {R.~A.}\ \bibnamefont {Porto}},\ }\href
  {https://doi.org/10.1103/PhysRevD.96.084065} {\bibfield  {journal} {\bibinfo
  {journal} {Phys. Rev. D}\ }\textbf {\bibinfo {volume} {96}},\ \bibinfo
  {pages} {084065} (\bibinfo {year} {2017}{\natexlab{a}})},\ \Eprint
  {https://arxiv.org/abs/1705.07938} {arXiv:1705.07938 [gr-qc]} \BibitemShut
  {NoStop}%
\bibitem [{\citenamefont {Maia}\ \emph
  {et~al.}(2017{\natexlab{b}})\citenamefont {Maia}, \citenamefont {Galley},
  \citenamefont {Leibovich},\ and\ \citenamefont {Porto}}]{Maia:2017gxn}%
  \BibitemOpen
  \bibfield  {author} {\bibinfo {author} {\bibfnamefont {N.~T.}\ \bibnamefont
  {Maia}}, \bibinfo {author} {\bibfnamefont {C.~R.}\ \bibnamefont {Galley}},
  \bibinfo {author} {\bibfnamefont {A.~K.}\ \bibnamefont {Leibovich}},\ and\
  \bibinfo {author} {\bibfnamefont {R.~A.}\ \bibnamefont {Porto}},\ }\href
  {https://doi.org/10.1103/PhysRevD.96.084064} {\bibfield  {journal} {\bibinfo
  {journal} {Phys. Rev. D}\ }\textbf {\bibinfo {volume} {96}},\ \bibinfo
  {pages} {084064} (\bibinfo {year} {2017}{\natexlab{b}})},\ \Eprint
  {https://arxiv.org/abs/1705.07934} {arXiv:1705.07934 [gr-qc]} \BibitemShut
  {NoStop}%
\bibitem [{\citenamefont {Marchand}\ \emph {et~al.}(2018)\citenamefont
  {Marchand}, \citenamefont {Bernard}, \citenamefont {Blanchet},\ and\
  \citenamefont {Faye}}]{Marchand:2017pir}%
  \BibitemOpen
  \bibfield  {author} {\bibinfo {author} {\bibfnamefont {T.}~\bibnamefont
  {Marchand}}, \bibinfo {author} {\bibfnamefont {L.}~\bibnamefont {Bernard}},
  \bibinfo {author} {\bibfnamefont {L.}~\bibnamefont {Blanchet}},\ and\
  \bibinfo {author} {\bibfnamefont {G.}~\bibnamefont {Faye}},\ }\href
  {https://doi.org/10.1103/PhysRevD.97.044023} {\bibfield  {journal} {\bibinfo
  {journal} {Phys. Rev. D}\ }\textbf {\bibinfo {volume} {97}},\ \bibinfo
  {pages} {044023} (\bibinfo {year} {2018})},\ \Eprint
  {https://arxiv.org/abs/1707.09289} {arXiv:1707.09289} \BibitemShut {NoStop}%
\bibitem [{\citenamefont {Foffa}\ and\ \citenamefont
  {Sturani}(2019)}]{Foffa:2019rdf}%
  \BibitemOpen
  \bibfield  {author} {\bibinfo {author} {\bibfnamefont {S.}~\bibnamefont
  {Foffa}}\ and\ \bibinfo {author} {\bibfnamefont {R.}~\bibnamefont
  {Sturani}},\ }\href {https://doi.org/10.1103/PhysRevD.100.024047} {\bibfield
  {journal} {\bibinfo  {journal} {Phys. Rev.}\ }\textbf {\bibinfo {volume}
  {D100}},\ \bibinfo {pages} {024047} (\bibinfo {year} {2019})},\ \Eprint
  {https://arxiv.org/abs/1903.05113} {arXiv:1903.05113 [gr-qc]} \BibitemShut
  {NoStop}%
\bibitem [{\citenamefont {Foffa}\ \emph
  {et~al.}(2019{\natexlab{a}})\citenamefont {Foffa}, \citenamefont {Porto},
  \citenamefont {Rothstein},\ and\ \citenamefont {Sturani}}]{nrgr4pn2}%
  \BibitemOpen
  \bibfield  {author} {\bibinfo {author} {\bibfnamefont {S.}~\bibnamefont
  {Foffa}}, \bibinfo {author} {\bibfnamefont {R.~A.}\ \bibnamefont {Porto}},
  \bibinfo {author} {\bibfnamefont {I.}~\bibnamefont {Rothstein}},\ and\
  \bibinfo {author} {\bibfnamefont {R.}~\bibnamefont {Sturani}},\ }\href
  {https://doi.org/10.1103/PhysRevD.100.024048} {\bibfield  {journal} {\bibinfo
   {journal} {Phys. Rev.}\ }\textbf {\bibinfo {volume} {D100}},\ \bibinfo
  {pages} {024048} (\bibinfo {year} {2019}{\natexlab{a}})},\ \Eprint
  {https://arxiv.org/abs/1903.05118} {arXiv:1903.05118} \BibitemShut {NoStop}%
\bibitem [{\citenamefont {Cho}\ \emph {et~al.}(2022{\natexlab{a}})\citenamefont
  {Cho}, \citenamefont {Porto},\ and\ \citenamefont {Yang}}]{Cho:2022syn}%
  \BibitemOpen
  \bibfield  {author} {\bibinfo {author} {\bibfnamefont {G.}~\bibnamefont
  {Cho}}, \bibinfo {author} {\bibfnamefont {R.~A.}\ \bibnamefont {Porto}},\
  and\ \bibinfo {author} {\bibfnamefont {Z.}~\bibnamefont {Yang}},\ }\href
  {https://doi.org/10.1103/PhysRevD.106.L101501} {\bibfield  {journal}
  {\bibinfo  {journal} {Phys. Rev. D}\ }\textbf {\bibinfo {volume} {106}},\
  \bibinfo {pages} {L101501} (\bibinfo {year} {2022}{\natexlab{a}})},\ \Eprint
  {https://arxiv.org/abs/2201.05138} {arXiv:2201.05138 [gr-qc]} \BibitemShut
  {NoStop}%
\bibitem [{\citenamefont {Blanchet}\ \emph {et~al.}(2023)\citenamefont
  {Blanchet}, \citenamefont {Faye}, \citenamefont {Henry}, \citenamefont
  {Larrouturou},\ and\ \citenamefont {Trestini}}]{Blanchet:2023sbv}%
  \BibitemOpen
  \bibfield  {author} {\bibinfo {author} {\bibfnamefont {L.}~\bibnamefont
  {Blanchet}}, \bibinfo {author} {\bibfnamefont {G.}~\bibnamefont {Faye}},
  \bibinfo {author} {\bibfnamefont {Q.}~\bibnamefont {Henry}}, \bibinfo
  {author} {\bibfnamefont {F.}~\bibnamefont {Larrouturou}},\ and\ \bibinfo
  {author} {\bibfnamefont {D.}~\bibnamefont {Trestini}},\ }\href
  {https://doi.org/10.1103/PhysRevD.108.064041} {\bibfield  {journal} {\bibinfo
   {journal} {Phys. Rev. D}\ }\textbf {\bibinfo {volume} {108}},\ \bibinfo
  {pages} {064041} (\bibinfo {year} {2023})},\ \Eprint
  {https://arxiv.org/abs/2304.11186} {arXiv:2304.11186 [gr-qc]} \BibitemShut
  {NoStop}%
\bibitem [{\citenamefont {Foffa}\ and\ \citenamefont {Sturani}(2020)}]{hered1}%
  \BibitemOpen
  \bibfield  {author} {\bibinfo {author} {\bibfnamefont {S.}~\bibnamefont
  {Foffa}}\ and\ \bibinfo {author} {\bibfnamefont {R.}~\bibnamefont
  {Sturani}},\ }\href {https://doi.org/10.1103/PhysRevD.101.064033} {\bibfield
  {journal} {\bibinfo  {journal} {Phys. Rev. D}\ }\textbf {\bibinfo {volume}
  {101}},\ \bibinfo {pages} {064033} (\bibinfo {year} {2020})},\ \Eprint
  {https://arxiv.org/abs/1907.02869} {arXiv:1907.02869 [gr-qc]} \BibitemShut
  {NoStop}%
\bibitem [{\citenamefont {Foffa}\ \emph
  {et~al.}(2019{\natexlab{b}})\citenamefont {Foffa}, \citenamefont {Mastrolia},
  \citenamefont {Sturani}, \citenamefont {Sturm},\ and\ \citenamefont
  {Torres~Bobadilla}}]{5pn1}%
  \BibitemOpen
  \bibfield  {author} {\bibinfo {author} {\bibfnamefont {S.}~\bibnamefont
  {Foffa}}, \bibinfo {author} {\bibfnamefont {P.}~\bibnamefont {Mastrolia}},
  \bibinfo {author} {\bibfnamefont {R.}~\bibnamefont {Sturani}}, \bibinfo
  {author} {\bibfnamefont {C.}~\bibnamefont {Sturm}},\ and\ \bibinfo {author}
  {\bibfnamefont {W.~J.}\ \bibnamefont {Torres~Bobadilla}},\ }\href
  {https://doi.org/10.1103/PhysRevLett.122.241605} {\bibfield  {journal}
  {\bibinfo  {journal} {Phys. Rev. Lett.}\ }\textbf {\bibinfo {volume} {122}},\
  \bibinfo {pages} {241605} (\bibinfo {year} {2019}{\natexlab{b}})},\ \Eprint
  {https://arxiv.org/abs/1902.10571} {arXiv:1902.10571} \BibitemShut {NoStop}%
\bibitem [{\citenamefont {Bl{\"u}mlein}\ \emph {et~al.}(2020)\citenamefont
  {Bl{\"u}mlein}, \citenamefont {Maier},\ and\ \citenamefont
  {Marquard}}]{5pn2}%
  \BibitemOpen
  \bibfield  {author} {\bibinfo {author} {\bibfnamefont {J.}~\bibnamefont
  {Bl{\"u}mlein}}, \bibinfo {author} {\bibfnamefont {A.}~\bibnamefont
  {Maier}},\ and\ \bibinfo {author} {\bibfnamefont {P.}~\bibnamefont
  {Marquard}},\ }\href@noop {} {\bibfield  {journal} {\bibinfo  {journal}
  {Phys. Lett. B}\ }\textbf {\bibinfo {volume} {800}},\ \bibinfo {pages}
  {135100} (\bibinfo {year} {2020})},\ \Eprint
  {https://arxiv.org/abs/1902.11180} {arXiv:1902.11180} \BibitemShut {NoStop}%
\bibitem [{\citenamefont {Almeida}\ \emph {et~al.}(2021)\citenamefont
  {Almeida}, \citenamefont {Foffa},\ and\ \citenamefont {Sturani}}]{hered2}%
  \BibitemOpen
  \bibfield  {author} {\bibinfo {author} {\bibfnamefont {G.~L.}\ \bibnamefont
  {Almeida}}, \bibinfo {author} {\bibfnamefont {S.}~\bibnamefont {Foffa}},\
  and\ \bibinfo {author} {\bibfnamefont {R.}~\bibnamefont {Sturani}},\ }\href
  {https://doi.org/10.1103/PhysRevD.104.124075} {\bibfield  {journal} {\bibinfo
   {journal} {Phys. Rev. D}\ }\textbf {\bibinfo {volume} {104}},\ \bibinfo
  {pages} {124075} (\bibinfo {year} {2021})},\ \Eprint
  {https://arxiv.org/abs/2110.14146} {arXiv:2110.14146 [gr-qc]} \BibitemShut
  {NoStop}%
\bibitem [{\citenamefont {Bl\"umlein}\ \emph
  {et~al.}(2021{\natexlab{a}})\citenamefont {Bl\"umlein}, \citenamefont
  {Maier}, \citenamefont {Marquard},\ and\ \citenamefont
  {Sch\"afer}}]{Blumlein:2020pyo}%
  \BibitemOpen
  \bibfield  {author} {\bibinfo {author} {\bibfnamefont {J.}~\bibnamefont
  {Bl\"umlein}}, \bibinfo {author} {\bibfnamefont {A.}~\bibnamefont {Maier}},
  \bibinfo {author} {\bibfnamefont {P.}~\bibnamefont {Marquard}},\ and\
  \bibinfo {author} {\bibfnamefont {G.}~\bibnamefont {Sch\"afer}},\ }\href
  {https://doi.org/10.1016/j.nuclphysb.2021.115352} {\bibfield  {journal}
  {\bibinfo  {journal} {Nucl. Phys. B}\ }\textbf {\bibinfo {volume} {965}},\
  \bibinfo {pages} {115352} (\bibinfo {year} {2021}{\natexlab{a}})},\ \Eprint
  {https://arxiv.org/abs/2010.13672} {arXiv:2010.13672 [gr-qc]} \BibitemShut
  {NoStop}%
\bibitem [{\citenamefont {Bini}\ \emph
  {et~al.}(2020{\natexlab{a}})\citenamefont {Bini}, \citenamefont {Damour},\
  and\ \citenamefont {Geralico}}]{Bini:2020wpo}%
  \BibitemOpen
  \bibfield  {author} {\bibinfo {author} {\bibfnamefont {D.}~\bibnamefont
  {Bini}}, \bibinfo {author} {\bibfnamefont {T.}~\bibnamefont {Damour}},\ and\
  \bibinfo {author} {\bibfnamefont {A.}~\bibnamefont {Geralico}},\ }\href
  {https://doi.org/10.1103/PhysRevD.102.024062} {\bibfield  {journal} {\bibinfo
   {journal} {Phys. Rev. D}\ }\textbf {\bibinfo {volume} {102}},\ \bibinfo
  {pages} {024062} (\bibinfo {year} {2020}{\natexlab{a}})},\ \Eprint
  {https://arxiv.org/abs/2003.11891} {arXiv:2003.11891 [gr-qc]} \BibitemShut
  {NoStop}%
\bibitem [{\citenamefont {Bini}\ \emph
  {et~al.}(2020{\natexlab{b}})\citenamefont {Bini}, \citenamefont {Damour},\
  and\ \citenamefont {Geralico}}]{binidam1}%
  \BibitemOpen
  \bibfield  {author} {\bibinfo {author} {\bibfnamefont {D.}~\bibnamefont
  {Bini}}, \bibinfo {author} {\bibfnamefont {T.}~\bibnamefont {Damour}},\ and\
  \bibinfo {author} {\bibfnamefont {A.}~\bibnamefont {Geralico}},\ }\href
  {https://doi.org/10.1103/PhysRevD.102.024061} {\bibfield  {journal} {\bibinfo
   {journal} {Phys. Rev. D}\ }\textbf {\bibinfo {volume} {102}},\ \bibinfo
  {pages} {024061} (\bibinfo {year} {2020}{\natexlab{b}})},\ \Eprint
  {https://arxiv.org/abs/2004.05407} {arXiv:2004.05407 [gr-qc]} \BibitemShut
  {NoStop}%
\bibitem [{\citenamefont {Bini}\ \emph
  {et~al.}(2020{\natexlab{c}})\citenamefont {Bini}, \citenamefont {Damour},\
  and\ \citenamefont {Geralico}}]{binidam2}%
  \BibitemOpen
  \bibfield  {author} {\bibinfo {author} {\bibfnamefont {D.}~\bibnamefont
  {Bini}}, \bibinfo {author} {\bibfnamefont {T.}~\bibnamefont {Damour}},\ and\
  \bibinfo {author} {\bibfnamefont {A.}~\bibnamefont {Geralico}},\ }\href
  {https://doi.org/10.1103/PhysRevD.102.084047} {\bibfield  {journal} {\bibinfo
   {journal} {Phys. Rev. D}\ }\textbf {\bibinfo {volume} {102}},\ \bibinfo
  {pages} {084047} (\bibinfo {year} {2020}{\natexlab{c}})},\ \Eprint
  {https://arxiv.org/abs/2007.11239} {arXiv:2007.11239 [gr-qc]} \BibitemShut
  {NoStop}%
\bibitem [{\citenamefont {Khalil}\ \emph {et~al.}(2022)\citenamefont {Khalil},
  \citenamefont {Buonanno}, \citenamefont {Steinhoff},\ and\ \citenamefont
  {Vines}}]{Khalil:2022ylj}%
  \BibitemOpen
  \bibfield  {author} {\bibinfo {author} {\bibfnamefont {M.}~\bibnamefont
  {Khalil}}, \bibinfo {author} {\bibfnamefont {A.}~\bibnamefont {Buonanno}},
  \bibinfo {author} {\bibfnamefont {J.}~\bibnamefont {Steinhoff}},\ and\
  \bibinfo {author} {\bibfnamefont {J.}~\bibnamefont {Vines}},\ }\href
  {https://doi.org/10.1103/PhysRevD.106.024042} {\bibfield  {journal} {\bibinfo
   {journal} {Phys. Rev. D}\ }\textbf {\bibinfo {volume} {106}},\ \bibinfo
  {pages} {024042} (\bibinfo {year} {2022})},\ \Eprint
  {https://arxiv.org/abs/2204.05047v3} {arXiv:2204.05047v3 [gr-qc]}
  \BibitemShut {NoStop}%
\bibitem [{\citenamefont {Bl\"umlein}\ \emph
  {et~al.}(2021{\natexlab{b}})\citenamefont {Bl\"umlein}, \citenamefont
  {Maier}, \citenamefont {Marquard},\ and\ \citenamefont
  {Sch\"afer}}]{Blumlein:2021txj}%
  \BibitemOpen
  \bibfield  {author} {\bibinfo {author} {\bibfnamefont {J.}~\bibnamefont
  {Bl\"umlein}}, \bibinfo {author} {\bibfnamefont {A.}~\bibnamefont {Maier}},
  \bibinfo {author} {\bibfnamefont {P.}~\bibnamefont {Marquard}},\ and\
  \bibinfo {author} {\bibfnamefont {G.}~\bibnamefont {Sch\"afer}},\ }\href
  {https://doi.org/10.1016/j.physletb.2021.136260} {\bibfield  {journal}
  {\bibinfo  {journal} {Phys. Lett. B}\ }\textbf {\bibinfo {volume} {816}},\
  \bibinfo {pages} {136260} (\bibinfo {year} {2021}{\natexlab{b}})},\ \Eprint
  {https://arxiv.org/abs/2101.08630} {arXiv:2101.08630 [gr-qc]} \BibitemShut
  {NoStop}%
\bibitem [{\citenamefont {Porto}\ \emph {et~al.}(2025)\citenamefont {Porto},
  \citenamefont {Riva},\ and\ \citenamefont {Yang}}]{memory}%
  \BibitemOpen
  \bibfield  {author} {\bibinfo {author} {\bibfnamefont {R.~A.}\ \bibnamefont
  {Porto}}, \bibinfo {author} {\bibfnamefont {M.~M.}\ \bibnamefont {Riva}},\
  and\ \bibinfo {author} {\bibfnamefont {Z.}~\bibnamefont {Yang}},\ }\href
  {https://doi.org/10.1007/JHEP04(2025)050} {\bibfield  {journal} {\bibinfo
  {journal} {JHEP}\ }\textbf {\bibinfo {volume} {04}},\ \bibinfo {pages}
  {050}},\ \Eprint {https://arxiv.org/abs/2409.05860} {arXiv:2409.05860
  [gr-qc]} \BibitemShut {NoStop}%
\bibitem [{\citenamefont {Almeida}\ \emph {et~al.}(2025)\citenamefont
  {Almeida}, \citenamefont {M\"uller}, \citenamefont {Foffa},\ and\
  \citenamefont {Sturani}}]{Almeida:2025nic}%
  \BibitemOpen
  \bibfield  {author} {\bibinfo {author} {\bibfnamefont {G.~L.}\ \bibnamefont
  {Almeida}}, \bibinfo {author} {\bibfnamefont {A.}~\bibnamefont {M\"uller}},
  \bibinfo {author} {\bibfnamefont {S.}~\bibnamefont {Foffa}},\ and\ \bibinfo
  {author} {\bibfnamefont {R.}~\bibnamefont {Sturani}},\ }\href@noop {} {\
  (\bibinfo {year} {2025})},\ \Eprint {https://arxiv.org/abs/2410.10565}
  {arXiv:2410.10565 [gr-qc]} \BibitemShut {NoStop}%
\bibitem [{\citenamefont {Cho}\ \emph {et~al.}(2021)\citenamefont {Cho},
  \citenamefont {Pardo},\ and\ \citenamefont {Porto}}]{Cho:2021mqw}%
  \BibitemOpen
  \bibfield  {author} {\bibinfo {author} {\bibfnamefont {G.}~\bibnamefont
  {Cho}}, \bibinfo {author} {\bibfnamefont {B.}~\bibnamefont {Pardo}},\ and\
  \bibinfo {author} {\bibfnamefont {R.~A.}\ \bibnamefont {Porto}},\ }\href
  {https://doi.org/10.1103/PhysRevD.104.024037} {\bibfield  {journal} {\bibinfo
   {journal} {Phys. Rev. D}\ }\textbf {\bibinfo {volume} {104}},\ \bibinfo
  {pages} {024037} (\bibinfo {year} {2021})},\ \Eprint
  {https://arxiv.org/abs/2103.14612} {arXiv:2103.14612 [gr-qc]} \BibitemShut
  {NoStop}%
\bibitem [{\citenamefont {Amalberti}\ \emph {et~al.}(2024)\citenamefont
  {Amalberti}, \citenamefont {Yang},\ and\ \citenamefont
  {Porto}}]{Amalberti:2024jaa}%
  \BibitemOpen
  \bibfield  {author} {\bibinfo {author} {\bibfnamefont {L.}~\bibnamefont
  {Amalberti}}, \bibinfo {author} {\bibfnamefont {Z.}~\bibnamefont {Yang}},\
  and\ \bibinfo {author} {\bibfnamefont {R.~A.}\ \bibnamefont {Porto}},\ }\href
  {https://doi.org/10.1103/PhysRevD.110.044046} {\bibfield  {journal} {\bibinfo
   {journal} {Phys. Rev. D}\ }\textbf {\bibinfo {volume} {110}},\ \bibinfo
  {pages} {044046} (\bibinfo {year} {2024})},\ \Eprint
  {https://arxiv.org/abs/2406.03457} {arXiv:2406.03457 [gr-qc]} \BibitemShut
  {NoStop}%
\bibitem [{\citenamefont {Bini}\ \emph {et~al.}(2024)\citenamefont {Bini},
  \citenamefont {Damour}, \citenamefont {De~Angelis}, \citenamefont {Geralico},
  \citenamefont {Herderschee}, \citenamefont {Roiban},\ and\ \citenamefont
  {Teng}}]{Bini:2024rsy}%
  \BibitemOpen
  \bibfield  {author} {\bibinfo {author} {\bibfnamefont {D.}~\bibnamefont
  {Bini}}, \bibinfo {author} {\bibfnamefont {T.}~\bibnamefont {Damour}},
  \bibinfo {author} {\bibfnamefont {S.}~\bibnamefont {De~Angelis}}, \bibinfo
  {author} {\bibfnamefont {A.}~\bibnamefont {Geralico}}, \bibinfo {author}
  {\bibfnamefont {A.}~\bibnamefont {Herderschee}}, \bibinfo {author}
  {\bibfnamefont {R.}~\bibnamefont {Roiban}},\ and\ \bibinfo {author}
  {\bibfnamefont {F.}~\bibnamefont {Teng}},\ }\href
  {https://doi.org/10.1103/PhysRevD.109.125008} {\bibfield  {journal} {\bibinfo
   {journal} {Phys. Rev. D}\ }\textbf {\bibinfo {volume} {109}},\ \bibinfo
  {pages} {125008} (\bibinfo {year} {2024})},\ \Eprint
  {https://arxiv.org/abs/2402.06604} {arXiv:2402.06604 [hep-th]} \BibitemShut
  {NoStop}%
\bibitem [{\citenamefont {Damour}(2016)}]{damour1}%
  \BibitemOpen
  \bibfield  {author} {\bibinfo {author} {\bibfnamefont {T.}~\bibnamefont
  {Damour}},\ }\href {https://doi.org/10.1103/PhysRevD.94.104015} {\bibfield
  {journal} {\bibinfo  {journal} {Phys. Rev.}\ }\textbf {\bibinfo {volume}
  {D94}},\ \bibinfo {pages} {104015} (\bibinfo {year} {2016})},\ \Eprint
  {https://arxiv.org/abs/1609.00354} {arXiv:1609.00354} \BibitemShut {NoStop}%
\bibitem [{\citenamefont {Damour}(2018)}]{Damour:2017zjx}%
  \BibitemOpen
  \bibfield  {author} {\bibinfo {author} {\bibfnamefont {T.}~\bibnamefont
  {Damour}},\ }\href {https://doi.org/10.1103/PhysRevD.97.044038} {\bibfield
  {journal} {\bibinfo  {journal} {Phys. Rev. D}\ }\textbf {\bibinfo {volume}
  {97}},\ \bibinfo {pages} {044038} (\bibinfo {year} {2018})},\ \Eprint
  {https://arxiv.org/abs/1710.10599} {arXiv:1710.10599 [gr-qc]} \BibitemShut
  {NoStop}%
\bibitem [{\citenamefont {Bjerrum-Bohr}\ \emph {et~al.}(2018)\citenamefont
  {Bjerrum-Bohr}, \citenamefont {Damgaard}, \citenamefont {Festuccia},
  \citenamefont {Plante},\ and\ \citenamefont {Vanhove}}]{bohr}%
  \BibitemOpen
  \bibfield  {author} {\bibinfo {author} {\bibfnamefont {N.~E.~J.}\
  \bibnamefont {Bjerrum-Bohr}}, \bibinfo {author} {\bibfnamefont {P.~H.}\
  \bibnamefont {Damgaard}}, \bibinfo {author} {\bibfnamefont {G.}~\bibnamefont
  {Festuccia}}, \bibinfo {author} {\bibfnamefont {L.}~\bibnamefont {Plante}},\
  and\ \bibinfo {author} {\bibfnamefont {P.}~\bibnamefont {Vanhove}},\ }\href
  {https://doi.org/10.1103/PhysRevLett.121.171601} {\bibfield  {journal}
  {\bibinfo  {journal} {Phys. Rev. Lett.}\ }\textbf {\bibinfo {volume} {121}},\
  \bibinfo {pages} {171601} (\bibinfo {year} {2018})},\ \Eprint
  {https://arxiv.org/abs/1806.04920} {arXiv:1806.04920} \BibitemShut {NoStop}%
\bibitem [{\citenamefont {Cheung}\ \emph {et~al.}(2018)\citenamefont {Cheung},
  \citenamefont {Rothstein},\ and\ \citenamefont {Solon}}]{cheung}%
  \BibitemOpen
  \bibfield  {author} {\bibinfo {author} {\bibfnamefont {C.}~\bibnamefont
  {Cheung}}, \bibinfo {author} {\bibfnamefont {I.~Z.}\ \bibnamefont
  {Rothstein}},\ and\ \bibinfo {author} {\bibfnamefont {M.~P.}\ \bibnamefont
  {Solon}},\ }\href {https://doi.org/10.1103/PhysRevLett.121.251101} {\bibfield
   {journal} {\bibinfo  {journal} {Phys. Rev. Lett.}\ }\textbf {\bibinfo
  {volume} {121}},\ \bibinfo {pages} {251101} (\bibinfo {year} {2018})},\
  \Eprint {https://arxiv.org/abs/1808.02489} {arXiv:1808.02489} \BibitemShut
  {NoStop}%
\bibitem [{\citenamefont {Kosower}\ \emph {et~al.}(2019)\citenamefont
  {Kosower}, \citenamefont {Maybee},\ and\ \citenamefont {O'Connell}}]{donal}%
  \BibitemOpen
  \bibfield  {author} {\bibinfo {author} {\bibfnamefont {D.~A.}\ \bibnamefont
  {Kosower}}, \bibinfo {author} {\bibfnamefont {B.}~\bibnamefont {Maybee}},\
  and\ \bibinfo {author} {\bibfnamefont {D.}~\bibnamefont {O'Connell}},\ }\href
  {https://doi.org/10.1007/JHEP02(2019)137} {\bibfield  {journal} {\bibinfo
  {journal} {JHEP}\ }\textbf {\bibinfo {volume} {02}},\ \bibinfo {pages}
  {137}},\ \Eprint {https://arxiv.org/abs/1811.10950} {arXiv:1811.10950}
  \BibitemShut {NoStop}%
\bibitem [{\citenamefont {Bern}\ \emph {et~al.}(2019)\citenamefont {Bern},
  \citenamefont {Cheung}, \citenamefont {Roiban}, \citenamefont {Shen},
  \citenamefont {Solon},\ and\ \citenamefont {Zeng}}]{zvi1}%
  \BibitemOpen
  \bibfield  {author} {\bibinfo {author} {\bibfnamefont {Z.}~\bibnamefont
  {Bern}}, \bibinfo {author} {\bibfnamefont {C.}~\bibnamefont {Cheung}},
  \bibinfo {author} {\bibfnamefont {R.}~\bibnamefont {Roiban}}, \bibinfo
  {author} {\bibfnamefont {C.-H.}\ \bibnamefont {Shen}}, \bibinfo {author}
  {\bibfnamefont {M.~P.}\ \bibnamefont {Solon}},\ and\ \bibinfo {author}
  {\bibfnamefont {M.}~\bibnamefont {Zeng}},\ }\href
  {https://doi.org/10.1103/PhysRevLett.122.201603} {\bibfield  {journal}
  {\bibinfo  {journal} {Phys. Rev. Lett.}\ }\textbf {\bibinfo {volume} {122}},\
  \bibinfo {pages} {201603} (\bibinfo {year} {2019})},\ \Eprint
  {https://arxiv.org/abs/1901.04424} {arXiv:1901.04424} \BibitemShut {NoStop}%
\bibitem [{\citenamefont {K{\"a}lin}\ and\ \citenamefont
  {Porto}(2020{\natexlab{a}})}]{paper1}%
  \BibitemOpen
  \bibfield  {author} {\bibinfo {author} {\bibfnamefont {G.}~\bibnamefont
  {K{\"a}lin}}\ and\ \bibinfo {author} {\bibfnamefont {R.~A.}\ \bibnamefont
  {Porto}},\ }\href {https://doi.org/10.1007/JHEP01(2020)072} {\bibfield
  {journal} {\bibinfo  {journal} {JHEP}\ }\textbf {\bibinfo {volume} {01}},\
  \bibinfo {pages} {072}},\ \Eprint {https://arxiv.org/abs/1910.03008}
  {arXiv:1910.03008} \BibitemShut {NoStop}%
\bibitem [{\citenamefont {K{\"a}lin}\ and\ \citenamefont
  {Porto}(2020{\natexlab{b}})}]{paper2}%
  \BibitemOpen
  \bibfield  {author} {\bibinfo {author} {\bibfnamefont {G.}~\bibnamefont
  {K{\"a}lin}}\ and\ \bibinfo {author} {\bibfnamefont {R.~A.}\ \bibnamefont
  {Porto}},\ }\href {https://doi.org/10.1007/JHEP02(2020)120} {\bibfield
  {journal} {\bibinfo  {journal} {JHEP}\ }\textbf {\bibinfo {volume} {02}},\
  \bibinfo {pages} {120}},\ \Eprint {https://arxiv.org/abs/1911.09130}
  {arXiv:1911.09130} \BibitemShut {NoStop}%
\bibitem [{\citenamefont {Cho}\ \emph {et~al.}(2022{\natexlab{b}})\citenamefont
  {Cho}, \citenamefont {K\"alin},\ and\ \citenamefont {Porto}}]{b2b3}%
  \BibitemOpen
  \bibfield  {author} {\bibinfo {author} {\bibfnamefont {G.}~\bibnamefont
  {Cho}}, \bibinfo {author} {\bibfnamefont {G.}~\bibnamefont {K\"alin}},\ and\
  \bibinfo {author} {\bibfnamefont {R.~A.}\ \bibnamefont {Porto}},\ }\href
  {https://doi.org/10.1007/JHEP04(2022)154} {\bibfield  {journal} {\bibinfo
  {journal} {JHEP}\ }\textbf {\bibinfo {volume} {04}},\ \bibinfo {pages}
  {154}},\ \bibinfo {note} {[Erratum: JHEP 07, 002 (2022)]},\ \Eprint
  {https://arxiv.org/abs/2112.03976} {arXiv:2112.03976 [hep-th]} \BibitemShut
  {NoStop}%
\bibitem [{\citenamefont {Damour}(2020)}]{Damour:2019lcq}%
  \BibitemOpen
  \bibfield  {author} {\bibinfo {author} {\bibfnamefont {T.}~\bibnamefont
  {Damour}},\ }\href {https://doi.org/10.1103/PhysRevD.102.024060} {\bibfield
  {journal} {\bibinfo  {journal} {Phys. Rev. D}\ }\textbf {\bibinfo {volume}
  {102}},\ \bibinfo {pages} {024060} (\bibinfo {year} {2020})},\ \Eprint
  {https://arxiv.org/abs/1912.02139} {arXiv:1912.02139 [gr-qc]} \BibitemShut
  {NoStop}%
\bibitem [{\citenamefont {K\"alin}\ and\ \citenamefont {Porto}(2020)}]{pmeft}%
  \BibitemOpen
  \bibfield  {author} {\bibinfo {author} {\bibfnamefont {G.}~\bibnamefont
  {K\"alin}}\ and\ \bibinfo {author} {\bibfnamefont {R.~A.}\ \bibnamefont
  {Porto}},\ }\href {https://doi.org/10.1007/JHEP11(2020)106} {\bibfield
  {journal} {\bibinfo  {journal} {JHEP}\ }\textbf {\bibinfo {volume} {11}},\
  \bibinfo {pages} {106}},\ \Eprint {https://arxiv.org/abs/2006.01184}
  {arXiv:2006.01184 [hep-th]} \BibitemShut {NoStop}%
\bibitem [{\citenamefont {K\"alin}\ \emph
  {et~al.}(2020{\natexlab{a}})\citenamefont {K\"alin}, \citenamefont {Liu},\
  and\ \citenamefont {Porto}}]{3pmeft}%
  \BibitemOpen
  \bibfield  {author} {\bibinfo {author} {\bibfnamefont {G.}~\bibnamefont
  {K\"alin}}, \bibinfo {author} {\bibfnamefont {Z.}~\bibnamefont {Liu}},\ and\
  \bibinfo {author} {\bibfnamefont {R.~A.}\ \bibnamefont {Porto}},\ }\href
  {https://doi.org/10.1103/PhysRevLett.125.261103} {\bibfield  {journal}
  {\bibinfo  {journal} {Phys. Rev. Lett.}\ }\textbf {\bibinfo {volume} {125}},\
  \bibinfo {pages} {261103} (\bibinfo {year} {2020}{\natexlab{a}})},\ \Eprint
  {https://arxiv.org/abs/2007.04977} {arXiv:2007.04977 [hep-th]} \BibitemShut
  {NoStop}%
\bibitem [{\citenamefont {K\"alin}\ \emph
  {et~al.}(2020{\natexlab{b}})\citenamefont {K\"alin}, \citenamefont {Liu},\
  and\ \citenamefont {Porto}}]{tidaleft}%
  \BibitemOpen
  \bibfield  {author} {\bibinfo {author} {\bibfnamefont {G.}~\bibnamefont
  {K\"alin}}, \bibinfo {author} {\bibfnamefont {Z.}~\bibnamefont {Liu}},\ and\
  \bibinfo {author} {\bibfnamefont {R.~A.}\ \bibnamefont {Porto}},\ }\href
  {https://doi.org/10.1103/PhysRevD.102.124025} {\bibfield  {journal} {\bibinfo
   {journal} {Phys. Rev. D}\ }\textbf {\bibinfo {volume} {102}},\ \bibinfo
  {pages} {124025} (\bibinfo {year} {2020}{\natexlab{b}})},\ \Eprint
  {https://arxiv.org/abs/2008.06047} {arXiv:2008.06047 [hep-th]} \BibitemShut
  {NoStop}%
\bibitem [{\citenamefont {Liu}\ \emph {et~al.}(2021)\citenamefont {Liu},
  \citenamefont {Porto},\ and\ \citenamefont {Yang}}]{pmefts}%
  \BibitemOpen
  \bibfield  {author} {\bibinfo {author} {\bibfnamefont {Z.}~\bibnamefont
  {Liu}}, \bibinfo {author} {\bibfnamefont {R.~A.}\ \bibnamefont {Porto}},\
  and\ \bibinfo {author} {\bibfnamefont {Z.}~\bibnamefont {Yang}},\ }\href
  {https://doi.org/10.1007/JHEP06(2021)012} {\bibfield  {journal} {\bibinfo
  {journal} {JHEP}\ }\textbf {\bibinfo {volume} {06}},\ \bibinfo {pages}
  {012}},\ \Eprint {https://arxiv.org/abs/2102.10059} {arXiv:2102.10059
  [hep-th]} \BibitemShut {NoStop}%
\bibitem [{\citenamefont {Jakobsen}\ \emph
  {et~al.}(2022{\natexlab{a}})\citenamefont {Jakobsen}, \citenamefont {Mogull},
  \citenamefont {Plefka},\ and\ \citenamefont {Steinhoff}}]{Jakobsen:2021zvh}%
  \BibitemOpen
  \bibfield  {author} {\bibinfo {author} {\bibfnamefont {G.~U.}\ \bibnamefont
  {Jakobsen}}, \bibinfo {author} {\bibfnamefont {G.}~\bibnamefont {Mogull}},
  \bibinfo {author} {\bibfnamefont {J.}~\bibnamefont {Plefka}},\ and\ \bibinfo
  {author} {\bibfnamefont {J.}~\bibnamefont {Steinhoff}},\ }\href
  {https://doi.org/10.1007/JHEP01(2022)027} {\bibfield  {journal} {\bibinfo
  {journal} {JHEP}\ }\textbf {\bibinfo {volume} {01}},\ \bibinfo {pages}
  {027}},\ \Eprint {https://arxiv.org/abs/2109.04465} {arXiv:2109.04465
  [hep-th]} \BibitemShut {NoStop}%
\bibitem [{\citenamefont {Mougiakakos}\ \emph {et~al.}(2021)\citenamefont
  {Mougiakakos}, \citenamefont {Riva},\ and\ \citenamefont
  {Vernizzi}}]{Mougiakakos:2021ckm}%
  \BibitemOpen
  \bibfield  {author} {\bibinfo {author} {\bibfnamefont {S.}~\bibnamefont
  {Mougiakakos}}, \bibinfo {author} {\bibfnamefont {M.~M.}\ \bibnamefont
  {Riva}},\ and\ \bibinfo {author} {\bibfnamefont {F.}~\bibnamefont
  {Vernizzi}},\ }\href {https://doi.org/10.1103/PhysRevD.104.024041} {\bibfield
   {journal} {\bibinfo  {journal} {Phys. Rev. D}\ }\textbf {\bibinfo {volume}
  {104}},\ \bibinfo {pages} {024041} (\bibinfo {year} {2021})},\ \Eprint
  {https://arxiv.org/abs/2102.08339} {arXiv:2102.08339 [gr-qc]} \BibitemShut
  {NoStop}%
\bibitem [{\citenamefont {Di~Vecchia}\ \emph {et~al.}(2021)\citenamefont
  {Di~Vecchia}, \citenamefont {Heissenberg}, \citenamefont {Russo},\ and\
  \citenamefont {Veneziano}}]{Gabriele2}%
  \BibitemOpen
  \bibfield  {author} {\bibinfo {author} {\bibfnamefont {P.}~\bibnamefont
  {Di~Vecchia}}, \bibinfo {author} {\bibfnamefont {C.}~\bibnamefont
  {Heissenberg}}, \bibinfo {author} {\bibfnamefont {R.}~\bibnamefont {Russo}},\
  and\ \bibinfo {author} {\bibfnamefont {G.}~\bibnamefont {Veneziano}},\ }\href
  {https://doi.org/10.1007/JHEP07(2021)169} {\bibfield  {journal} {\bibinfo
  {journal} {JHEP}\ }\textbf {\bibinfo {volume} {07}},\ \bibinfo {pages}
  {169}},\ \Eprint {https://arxiv.org/abs/2104.03256} {arXiv:2104.03256
  [hep-th]} \BibitemShut {NoStop}%
\bibitem [{\citenamefont {K\"alin}\ \emph {et~al.}(2023)\citenamefont
  {K\"alin}, \citenamefont {Neef},\ and\ \citenamefont {Porto}}]{eftrad}%
  \BibitemOpen
  \bibfield  {author} {\bibinfo {author} {\bibfnamefont {G.}~\bibnamefont
  {K\"alin}}, \bibinfo {author} {\bibfnamefont {J.}~\bibnamefont {Neef}},\ and\
  \bibinfo {author} {\bibfnamefont {R.~A.}\ \bibnamefont {Porto}},\ }\href
  {https://doi.org/10.1007/JHEP01(2023)140} {\bibfield  {journal} {\bibinfo
  {journal} {JHEP}\ }\textbf {\bibinfo {volume} {01}},\ \bibinfo {pages}
  {140}},\ \Eprint {https://arxiv.org/abs/2207.00580} {arXiv:2207.00580
  [hep-th]} \BibitemShut {NoStop}%
\bibitem [{\citenamefont {Jakobsen}\ \emph
  {et~al.}(2022{\natexlab{b}})\citenamefont {Jakobsen}, \citenamefont {Mogull},
  \citenamefont {Plefka},\ and\ \citenamefont {Sauer}}]{Jakobsen:2022psy}%
  \BibitemOpen
  \bibfield  {author} {\bibinfo {author} {\bibfnamefont {G.~U.}\ \bibnamefont
  {Jakobsen}}, \bibinfo {author} {\bibfnamefont {G.}~\bibnamefont {Mogull}},
  \bibinfo {author} {\bibfnamefont {J.}~\bibnamefont {Plefka}},\ and\ \bibinfo
  {author} {\bibfnamefont {B.}~\bibnamefont {Sauer}},\ }\href
  {https://doi.org/10.1007/JHEP10(2022)128} {\bibfield  {journal} {\bibinfo
  {journal} {JHEP}\ }\textbf {\bibinfo {volume} {10}},\ \bibinfo {pages}
  {128}},\ \Eprint {https://arxiv.org/abs/2207.00569} {arXiv:2207.00569
  [hep-th]} \BibitemShut {NoStop}%
\bibitem [{\citenamefont {Dlapa}\ \emph
  {et~al.}(2022{\natexlab{a}})\citenamefont {Dlapa}, \citenamefont {K\"alin},
  \citenamefont {Liu},\ and\ \citenamefont {Porto}}]{4pmeft}%
  \BibitemOpen
  \bibfield  {author} {\bibinfo {author} {\bibfnamefont {C.}~\bibnamefont
  {Dlapa}}, \bibinfo {author} {\bibfnamefont {G.}~\bibnamefont {K\"alin}},
  \bibinfo {author} {\bibfnamefont {Z.}~\bibnamefont {Liu}},\ and\ \bibinfo
  {author} {\bibfnamefont {R.~A.}\ \bibnamefont {Porto}},\ }\href
  {https://doi.org/10.1016/j.physletb.2022.137203} {\bibfield  {journal}
  {\bibinfo  {journal} {Phys. Lett. B}\ }\textbf {\bibinfo {volume} {831}},\
  \bibinfo {pages} {137203} (\bibinfo {year} {2022}{\natexlab{a}})},\ \Eprint
  {https://arxiv.org/abs/2106.08276} {2106.08276 [hep-th]} \BibitemShut
  {NoStop}%
\bibitem [{\citenamefont {Dlapa}\ \emph
  {et~al.}(2022{\natexlab{b}})\citenamefont {Dlapa}, \citenamefont {K\"alin},
  \citenamefont {Liu},\ and\ \citenamefont {Porto}}]{4pmeft2}%
  \BibitemOpen
  \bibfield  {author} {\bibinfo {author} {\bibfnamefont {C.}~\bibnamefont
  {Dlapa}}, \bibinfo {author} {\bibfnamefont {G.}~\bibnamefont {K\"alin}},
  \bibinfo {author} {\bibfnamefont {Z.}~\bibnamefont {Liu}},\ and\ \bibinfo
  {author} {\bibfnamefont {R.~A.}\ \bibnamefont {Porto}},\ }\href
  {https://doi.org/10.1103/PhysRevLett.128.161104} {\bibfield  {journal}
  {\bibinfo  {journal} {Phys. Rev. Lett.}\ }\textbf {\bibinfo {volume} {128}},\
  \bibinfo {pages} {161104} (\bibinfo {year} {2022}{\natexlab{b}})},\ \Eprint
  {https://arxiv.org/abs/2112.11296} {arXiv:2112.11296 [hep-th]} \BibitemShut
  {NoStop}%
\bibitem [{\citenamefont {Dlapa}\ \emph
  {et~al.}(2023{\natexlab{a}})\citenamefont {Dlapa}, \citenamefont {K{\"a}lin},
  \citenamefont {Liu}, \citenamefont {Neef},\ and\ \citenamefont
  {Porto}}]{4pmeftot}%
  \BibitemOpen
  \bibfield  {author} {\bibinfo {author} {\bibfnamefont {C.}~\bibnamefont
  {Dlapa}}, \bibinfo {author} {\bibfnamefont {G.}~\bibnamefont {K{\"a}lin}},
  \bibinfo {author} {\bibfnamefont {Z.}~\bibnamefont {Liu}}, \bibinfo {author}
  {\bibfnamefont {J.}~\bibnamefont {Neef}},\ and\ \bibinfo {author}
  {\bibfnamefont {R.~A.}\ \bibnamefont {Porto}},\ }\href
  {https://doi.org/10.1103/PhysRevLett.130.101401} {\bibfield  {journal}
  {\bibinfo  {journal} {Phys. Rev. Lett.}\ }\textbf {\bibinfo {volume} {130}},\
  \bibinfo {pages} {101401} (\bibinfo {year} {2023}{\natexlab{a}})},\ \Eprint
  {https://arxiv.org/abs/2210.05541} {arXiv:2210.05541 [hep-th]} \BibitemShut
  {NoStop}%
\bibitem [{\citenamefont {Bern}\ \emph {et~al.}(2021)\citenamefont {Bern},
  \citenamefont {Parra-Martinez}, \citenamefont {Roiban}, \citenamefont {Ruf},
  \citenamefont {Shen}, \citenamefont {Solon},\ and\ \citenamefont
  {Zeng}}]{4pmzvi}%
  \BibitemOpen
  \bibfield  {author} {\bibinfo {author} {\bibfnamefont {Z.}~\bibnamefont
  {Bern}}, \bibinfo {author} {\bibfnamefont {J.}~\bibnamefont
  {Parra-Martinez}}, \bibinfo {author} {\bibfnamefont {R.}~\bibnamefont
  {Roiban}}, \bibinfo {author} {\bibfnamefont {M.~S.}\ \bibnamefont {Ruf}},
  \bibinfo {author} {\bibfnamefont {C.-H.}\ \bibnamefont {Shen}}, \bibinfo
  {author} {\bibfnamefont {M.~P.}\ \bibnamefont {Solon}},\ and\ \bibinfo
  {author} {\bibfnamefont {M.}~\bibnamefont {Zeng}},\ }\href
  {https://doi.org/10.1103/PhysRevLett.126.171601} {\bibfield  {journal}
  {\bibinfo  {journal} {Phys. Rev. Lett.}\ }\textbf {\bibinfo {volume} {126}},\
  \bibinfo {pages} {171601} (\bibinfo {year} {2021})},\ \Eprint
  {https://arxiv.org/abs/2101.07254} {arXiv:2101.07254 [hep-th]} \BibitemShut
  {NoStop}%
\bibitem [{\citenamefont {Bern}\ \emph {et~al.}(2022)\citenamefont {Bern},
  \citenamefont {Parra-Martinez}, \citenamefont {Roiban}, \citenamefont {Ruf},
  \citenamefont {Shen}, \citenamefont {Solon},\ and\ \citenamefont
  {Zeng}}]{4pmzvi2}%
  \BibitemOpen
  \bibfield  {author} {\bibinfo {author} {\bibfnamefont {Z.}~\bibnamefont
  {Bern}}, \bibinfo {author} {\bibfnamefont {J.}~\bibnamefont
  {Parra-Martinez}}, \bibinfo {author} {\bibfnamefont {R.}~\bibnamefont
  {Roiban}}, \bibinfo {author} {\bibfnamefont {M.~S.}\ \bibnamefont {Ruf}},
  \bibinfo {author} {\bibfnamefont {C.-H.}\ \bibnamefont {Shen}}, \bibinfo
  {author} {\bibfnamefont {M.~P.}\ \bibnamefont {Solon}},\ and\ \bibinfo
  {author} {\bibfnamefont {M.}~\bibnamefont {Zeng}},\ }\href
  {https://doi.org/10.1103/PhysRevLett.128.161103} {\bibfield  {journal}
  {\bibinfo  {journal} {Phys. Rev. Lett.}\ }\textbf {\bibinfo {volume} {128}},\
  \bibinfo {pages} {161103} (\bibinfo {year} {2022})},\ \Eprint
  {https://arxiv.org/abs/2112.10750} {arXiv:2112.10750 [hep-th]} \BibitemShut
  {NoStop}%
\bibitem [{\citenamefont {Bini}\ \emph {et~al.}(2023)\citenamefont {Bini},
  \citenamefont {Damour},\ and\ \citenamefont {Geralico}}]{Bini:2022enm}%
  \BibitemOpen
  \bibfield  {author} {\bibinfo {author} {\bibfnamefont {D.}~\bibnamefont
  {Bini}}, \bibinfo {author} {\bibfnamefont {T.}~\bibnamefont {Damour}},\ and\
  \bibinfo {author} {\bibfnamefont {A.}~\bibnamefont {Geralico}},\ }\href
  {https://doi.org/10.1103/PhysRevD.107.024012} {\bibfield  {journal} {\bibinfo
   {journal} {Phys. Rev. D}\ }\textbf {\bibinfo {volume} {107}},\ \bibinfo
  {pages} {024012} (\bibinfo {year} {2023})},\ \Eprint
  {https://arxiv.org/abs/2210.07165} {arXiv:2210.07165 [gr-qc]} \BibitemShut
  {NoStop}%
\bibitem [{\citenamefont {Damgaard}\ \emph {et~al.}(2023)\citenamefont
  {Damgaard}, \citenamefont {Hansen}, \citenamefont {Plant\'e},\ and\
  \citenamefont {Vanhove}}]{Damgaard:2023ttc}%
  \BibitemOpen
  \bibfield  {author} {\bibinfo {author} {\bibfnamefont {P.~H.}\ \bibnamefont
  {Damgaard}}, \bibinfo {author} {\bibfnamefont {E.~R.}\ \bibnamefont
  {Hansen}}, \bibinfo {author} {\bibfnamefont {L.}~\bibnamefont {Plant\'e}},\
  and\ \bibinfo {author} {\bibfnamefont {P.}~\bibnamefont {Vanhove}},\ }\href
  {https://doi.org/10.1007/JHEP09(2023)183} {\bibfield  {journal} {\bibinfo
  {journal} {JHEP}\ }\textbf {\bibinfo {volume} {09}},\ \bibinfo {pages}
  {183}},\ \Eprint {https://arxiv.org/abs/2307.04746} {arXiv:2307.04746
  [hep-th]} \BibitemShut {NoStop}%
\bibitem [{\citenamefont {Dlapa}\ \emph
  {et~al.}(2023{\natexlab{b}})\citenamefont {Dlapa}, \citenamefont {K\"alin},
  \citenamefont {Liu},\ and\ \citenamefont {Porto}}]{dklp}%
  \BibitemOpen
  \bibfield  {author} {\bibinfo {author} {\bibfnamefont {C.}~\bibnamefont
  {Dlapa}}, \bibinfo {author} {\bibfnamefont {G.}~\bibnamefont {K\"alin}},
  \bibinfo {author} {\bibfnamefont {Z.}~\bibnamefont {Liu}},\ and\ \bibinfo
  {author} {\bibfnamefont {R.~A.}\ \bibnamefont {Porto}},\ }\href
  {https://doi.org/10.1007/JHEP08(2023)109} {\bibfield  {journal} {\bibinfo
  {journal} {JHEP}\ }\textbf {\bibinfo {volume} {08}},\ \bibinfo {pages}
  {109}},\ \Eprint {https://arxiv.org/abs/2304.01275} {arXiv:2304.01275
  [hep-th]} \BibitemShut {NoStop}%
\bibitem [{\citenamefont {Klemm}\ \emph {et~al.}(2024)\citenamefont {Klemm},
  \citenamefont {Nega}, \citenamefont {Sauer},\ and\ \citenamefont
  {Plefka}}]{cy}%
  \BibitemOpen
  \bibfield  {author} {\bibinfo {author} {\bibfnamefont {A.}~\bibnamefont
  {Klemm}}, \bibinfo {author} {\bibfnamefont {C.}~\bibnamefont {Nega}},
  \bibinfo {author} {\bibfnamefont {B.}~\bibnamefont {Sauer}},\ and\ \bibinfo
  {author} {\bibfnamefont {J.}~\bibnamefont {Plefka}},\ }\href
  {https://doi.org/10.1103/PhysRevD.109.124046} {\bibfield  {journal} {\bibinfo
   {journal} {Phys. Rev. D}\ }\textbf {\bibinfo {volume} {109}},\ \bibinfo
  {pages} {124046} (\bibinfo {year} {2024})},\ \Eprint
  {https://arxiv.org/abs/2401.07899} {arXiv:2401.07899 [hep-th]} \BibitemShut
  {NoStop}%
\bibitem [{\citenamefont {Frellesvig}\ \emph {et~al.}(2024)\citenamefont
  {Frellesvig}, \citenamefont {Morales},\ and\ \citenamefont
  {Wilhelm}}]{Frellesvig:2023bbf}%
  \BibitemOpen
  \bibfield  {author} {\bibinfo {author} {\bibfnamefont {H.}~\bibnamefont
  {Frellesvig}}, \bibinfo {author} {\bibfnamefont {R.}~\bibnamefont
  {Morales}},\ and\ \bibinfo {author} {\bibfnamefont {M.}~\bibnamefont
  {Wilhelm}},\ }\href {https://doi.org/10.1103/PhysRevLett.132.201602}
  {\bibfield  {journal} {\bibinfo  {journal} {Phys. Rev. Lett.}\ }\textbf
  {\bibinfo {volume} {132}},\ \bibinfo {pages} {201602} (\bibinfo {year}
  {2024})},\ \Eprint {https://arxiv.org/abs/2312.11371} {arXiv:2312.11371
  [hep-th]} \BibitemShut {NoStop}%
\bibitem [{\citenamefont {Bern}\ \emph
  {et~al.}(2025{\natexlab{a}})\citenamefont {Bern}, \citenamefont {Herrmann},
  \citenamefont {Roiban}, \citenamefont {Ruf},\ and\ \citenamefont
  {Zeng}}]{Bern:2024vqs}%
  \BibitemOpen
  \bibfield  {author} {\bibinfo {author} {\bibfnamefont {Z.}~\bibnamefont
  {Bern}}, \bibinfo {author} {\bibfnamefont {E.}~\bibnamefont {Herrmann}},
  \bibinfo {author} {\bibfnamefont {R.}~\bibnamefont {Roiban}}, \bibinfo
  {author} {\bibfnamefont {M.~S.}\ \bibnamefont {Ruf}},\ and\ \bibinfo {author}
  {\bibfnamefont {M.}~\bibnamefont {Zeng}},\ }\href
  {https://doi.org/10.1007/JHEP06(2025)115} {\bibfield  {journal} {\bibinfo
  {journal} {JHEP}\ }\textbf {\bibinfo {volume} {06}},\ \bibinfo {pages}
  {115}},\ \Eprint {https://arxiv.org/abs/2408.06686} {arXiv:2408.06686
  [hep-th]} \BibitemShut {NoStop}%
\bibitem [{\citenamefont {Bern}\ \emph {et~al.}(2024)\citenamefont {Bern},
  \citenamefont {Herrmann}, \citenamefont {Roiban}, \citenamefont {Ruf},
  \citenamefont {Smirnov}, \citenamefont {Smirnov},\ and\ \citenamefont
  {Zeng}}]{Bern:2024adl}%
  \BibitemOpen
  \bibfield  {author} {\bibinfo {author} {\bibfnamefont {Z.}~\bibnamefont
  {Bern}}, \bibinfo {author} {\bibfnamefont {E.}~\bibnamefont {Herrmann}},
  \bibinfo {author} {\bibfnamefont {R.}~\bibnamefont {Roiban}}, \bibinfo
  {author} {\bibfnamefont {M.~S.}\ \bibnamefont {Ruf}}, \bibinfo {author}
  {\bibfnamefont {A.~V.}\ \bibnamefont {Smirnov}}, \bibinfo {author}
  {\bibfnamefont {V.~A.}\ \bibnamefont {Smirnov}},\ and\ \bibinfo {author}
  {\bibfnamefont {M.}~\bibnamefont {Zeng}},\ }\href
  {https://doi.org/10.1007/JHEP10(2024)023} {\bibfield  {journal} {\bibinfo
  {journal} {JHEP}\ }\textbf {\bibinfo {volume} {10}},\ \bibinfo {pages}
  {023}},\ \Eprint {https://arxiv.org/abs/2406.01554} {arXiv:2406.01554
  [hep-th]} \BibitemShut {NoStop}%
\bibitem [{\citenamefont {Buonanno}\ \emph {et~al.}(2024)\citenamefont
  {Buonanno}, \citenamefont {Jakobsen},\ and\ \citenamefont
  {Mogull}}]{Buonanno:2024vkx}%
  \BibitemOpen
  \bibfield  {author} {\bibinfo {author} {\bibfnamefont {A.}~\bibnamefont
  {Buonanno}}, \bibinfo {author} {\bibfnamefont {G.~U.}\ \bibnamefont
  {Jakobsen}},\ and\ \bibinfo {author} {\bibfnamefont {G.}~\bibnamefont
  {Mogull}},\ }\href {https://doi.org/10.1103/PhysRevD.110.044038} {\bibfield
  {journal} {\bibinfo  {journal} {Phys. Rev. D}\ }\textbf {\bibinfo {volume}
  {110}},\ \bibinfo {pages} {044038} (\bibinfo {year} {2024})},\ \Eprint
  {https://arxiv.org/abs/2402.12342} {arXiv:2402.12342 [gr-qc]} \BibitemShut
  {NoStop}%
\bibitem [{\citenamefont {Dlapa}\ \emph {et~al.}(2024)\citenamefont {Dlapa},
  \citenamefont {K\"alin}, \citenamefont {Liu},\ and\ \citenamefont
  {Porto}}]{4pmeftloc}%
  \BibitemOpen
  \bibfield  {author} {\bibinfo {author} {\bibfnamefont {C.}~\bibnamefont
  {Dlapa}}, \bibinfo {author} {\bibfnamefont {G.}~\bibnamefont {K\"alin}},
  \bibinfo {author} {\bibfnamefont {Z.}~\bibnamefont {Liu}},\ and\ \bibinfo
  {author} {\bibfnamefont {R.~A.}\ \bibnamefont {Porto}},\ }\href
  {https://doi.org/10.1103/PhysRevLett.132.221401} {\bibfield  {journal}
  {\bibinfo  {journal} {Phys. Rev. Lett.}\ }\textbf {\bibinfo {volume} {132}},\
  \bibinfo {pages} {221401} (\bibinfo {year} {2024})},\ \Eprint
  {https://arxiv.org/abs/2403.04853} {arXiv:2403.04853 [hep-th]} \BibitemShut
  {NoStop}%
\bibitem [{\citenamefont {Dlapa}\ \emph {et~al.}(2025)\citenamefont {Dlapa},
  \citenamefont {K{\"a}lin}, \citenamefont {Liu},\ and\ \citenamefont
  {Porto}}]{5pmeft1loc}%
  \BibitemOpen
  \bibfield  {author} {\bibinfo {author} {\bibfnamefont {C.}~\bibnamefont
  {Dlapa}}, \bibinfo {author} {\bibfnamefont {G.}~\bibnamefont {K{\"a}lin}},
  \bibinfo {author} {\bibfnamefont {Z.}~\bibnamefont {Liu}},\ and\ \bibinfo
  {author} {\bibfnamefont {R.~A.}\ \bibnamefont {Porto}},\ }\href
  {https://doi.org/10.1103/215k-27sj} {\bibfield  {journal} {\bibinfo
  {journal} {Phys. Rev. Lett.}\ }\textbf {\bibinfo {volume} {135}},\ \bibinfo
  {pages} {251401} (\bibinfo {year} {2025})},\ \Eprint
  {https://arxiv.org/abs/2506.20665} {arXiv:2506.20665 [hep-th]} \BibitemShut
  {NoStop}%
\bibitem [{\citenamefont {Cheung}\ \emph {et~al.}(2024)\citenamefont {Cheung},
  \citenamefont {Parra-Martinez}, \citenamefont {Rothstein}, \citenamefont
  {Shah},\ and\ \citenamefont {Wilson-Gerow}}]{Cheung:2024byb}%
  \BibitemOpen
  \bibfield  {author} {\bibinfo {author} {\bibfnamefont {C.}~\bibnamefont
  {Cheung}}, \bibinfo {author} {\bibfnamefont {J.}~\bibnamefont
  {Parra-Martinez}}, \bibinfo {author} {\bibfnamefont {I.~Z.}\ \bibnamefont
  {Rothstein}}, \bibinfo {author} {\bibfnamefont {N.}~\bibnamefont {Shah}},\
  and\ \bibinfo {author} {\bibfnamefont {J.}~\bibnamefont {Wilson-Gerow}},\
  }\href {https://doi.org/10.1007/JHEP10(2024)005} {\bibfield  {journal}
  {\bibinfo  {journal} {JHEP}\ }\textbf {\bibinfo {volume} {10}},\ \bibinfo
  {pages} {005}},\ \Eprint {https://arxiv.org/abs/2406.14770} {arXiv:2406.14770
  [hep-th]} \BibitemShut {NoStop}%
\bibitem [{\citenamefont {Bini}\ and\ \citenamefont
  {Damour}(2024)}]{Bini:2024tft}%
  \BibitemOpen
  \bibfield  {author} {\bibinfo {author} {\bibfnamefont {D.}~\bibnamefont
  {Bini}}\ and\ \bibinfo {author} {\bibfnamefont {T.}~\bibnamefont {Damour}},\
  }\href {https://doi.org/10.1103/PhysRevD.110.064005} {\bibfield  {journal}
  {\bibinfo  {journal} {Phys. Rev. D}\ }\textbf {\bibinfo {volume} {110}},\
  \bibinfo {pages} {064005} (\bibinfo {year} {2024})},\ \Eprint
  {https://arxiv.org/abs/2406.04878} {arXiv:2406.04878 [gr-qc]} \BibitemShut
  {NoStop}%
\bibitem [{\citenamefont {Driesse}\ \emph {et~al.}(2024)\citenamefont
  {Driesse}, \citenamefont {Jakobsen}, \citenamefont {Mogull}, \citenamefont
  {Plefka}, \citenamefont {Sauer},\ and\ \citenamefont
  {Usovitsch}}]{Driesse:2024xad}%
  \BibitemOpen
  \bibfield  {author} {\bibinfo {author} {\bibfnamefont {M.}~\bibnamefont
  {Driesse}}, \bibinfo {author} {\bibfnamefont {G.~U.}\ \bibnamefont
  {Jakobsen}}, \bibinfo {author} {\bibfnamefont {G.}~\bibnamefont {Mogull}},
  \bibinfo {author} {\bibfnamefont {J.}~\bibnamefont {Plefka}}, \bibinfo
  {author} {\bibfnamefont {B.}~\bibnamefont {Sauer}},\ and\ \bibinfo {author}
  {\bibfnamefont {J.}~\bibnamefont {Usovitsch}},\ }\href
  {https://doi.org/10.1103/PhysRevLett.132.241402} {\bibfield  {journal}
  {\bibinfo  {journal} {Phys. Rev. Lett.}\ }\textbf {\bibinfo {volume} {132}},\
  \bibinfo {pages} {241402} (\bibinfo {year} {2024})},\ \Eprint
  {https://arxiv.org/abs/2403.07781} {arXiv:2403.07781 [hep-th]} \BibitemShut
  {NoStop}%
\bibitem [{\citenamefont {Driesse}\ \emph {et~al.}(2025)\citenamefont
  {Driesse}, \citenamefont {Jakobsen}, \citenamefont {Klemm}, \citenamefont
  {Mogull}, \citenamefont {Nega}, \citenamefont {Plefka}, \citenamefont
  {Sauer},\ and\ \citenamefont {Usovitsch}}]{Driesse:2024feo}%
  \BibitemOpen
  \bibfield  {author} {\bibinfo {author} {\bibfnamefont {M.}~\bibnamefont
  {Driesse}}, \bibinfo {author} {\bibfnamefont {G.~U.}\ \bibnamefont
  {Jakobsen}}, \bibinfo {author} {\bibfnamefont {A.}~\bibnamefont {Klemm}},
  \bibinfo {author} {\bibfnamefont {G.}~\bibnamefont {Mogull}}, \bibinfo
  {author} {\bibfnamefont {C.}~\bibnamefont {Nega}}, \bibinfo {author}
  {\bibfnamefont {J.}~\bibnamefont {Plefka}}, \bibinfo {author} {\bibfnamefont
  {B.}~\bibnamefont {Sauer}},\ and\ \bibinfo {author} {\bibfnamefont
  {J.}~\bibnamefont {Usovitsch}},\ }\href
  {https://doi.org/10.1038/s41586-025-08984-2} {\bibfield  {journal} {\bibinfo
  {journal} {Nature}\ }\textbf {\bibinfo {volume} {641}},\ \bibinfo {pages}
  {603} (\bibinfo {year} {2025})},\ \Eprint {https://arxiv.org/abs/2411.11846}
  {arXiv:2411.11846 [hep-th]} \BibitemShut {NoStop}%
\bibitem [{\citenamefont {Heissenberg}(2025)}]{Heissenberg:2025ocy}%
  \BibitemOpen
  \bibfield  {author} {\bibinfo {author} {\bibfnamefont {C.}~\bibnamefont
  {Heissenberg}},\ }\href {https://doi.org/10.1103/xz14-87q7} {\bibfield
  {journal} {\bibinfo  {journal} {Phys. Rev. D}\ }\textbf {\bibinfo {volume}
  {111}},\ \bibinfo {pages} {126012} (\bibinfo {year} {2025})},\ \Eprint
  {https://arxiv.org/abs/2501.02904} {arXiv:2501.02904 [hep-th]} \BibitemShut
  {NoStop}%
\bibitem [{\citenamefont {Heissenberg}\ and\ \citenamefont
  {Russo}(2025)}]{Heissenberg:2025fcr}%
  \BibitemOpen
  \bibfield  {author} {\bibinfo {author} {\bibfnamefont {C.}~\bibnamefont
  {Heissenberg}}\ and\ \bibinfo {author} {\bibfnamefont {R.}~\bibnamefont
  {Russo}},\ }\href@noop {} {\  (\bibinfo {year} {2025})},\ \Eprint
  {https://arxiv.org/abs/2511.13835} {arXiv:2511.13835 [hep-th]} \BibitemShut
  {NoStop}%
\bibitem [{\citenamefont {Bini}\ and\ \citenamefont
  {Damour}(2025)}]{Bini:2025vuk}%
  \BibitemOpen
  \bibfield  {author} {\bibinfo {author} {\bibfnamefont {D.}~\bibnamefont
  {Bini}}\ and\ \bibinfo {author} {\bibfnamefont {T.}~\bibnamefont {Damour}},\
  }\href {https://doi.org/10.1103/8ks7-2blq} {\bibfield  {journal} {\bibinfo
  {journal} {Phys. Rev. D}\ }\textbf {\bibinfo {volume} {112}},\ \bibinfo
  {pages} {044002} (\bibinfo {year} {2025})},\ \Eprint
  {https://arxiv.org/abs/2504.20204} {arXiv:2504.20204 [hep-th]} \BibitemShut
  {NoStop}%
\bibitem [{\citenamefont {Bern}\ \emph
  {et~al.}(2025{\natexlab{b}})\citenamefont {Bern}, \citenamefont {Herrmann},
  \citenamefont {Roiban}, \citenamefont {Ruf}, \citenamefont {Smirnov},
  \citenamefont {Smith},\ and\ \citenamefont {Zeng}}]{Bern:2025wyd}%
  \BibitemOpen
  \bibfield  {author} {\bibinfo {author} {\bibfnamefont {Z.}~\bibnamefont
  {Bern}}, \bibinfo {author} {\bibfnamefont {E.}~\bibnamefont {Herrmann}},
  \bibinfo {author} {\bibfnamefont {R.}~\bibnamefont {Roiban}}, \bibinfo
  {author} {\bibfnamefont {M.~S.}\ \bibnamefont {Ruf}}, \bibinfo {author}
  {\bibfnamefont {A.~V.}\ \bibnamefont {Smirnov}}, \bibinfo {author}
  {\bibfnamefont {S.}~\bibnamefont {Smith}},\ and\ \bibinfo {author}
  {\bibfnamefont {M.}~\bibnamefont {Zeng}},\ }\href@noop {} {\  (\bibinfo
  {year} {2025}{\natexlab{b}})},\ \Eprint {https://arxiv.org/abs/2512.23654}
  {arXiv:2512.23654 [hep-th]} \BibitemShut {NoStop}%
\bibitem [{\citenamefont {Blanchet}\ \emph {et~al.}(2026)\citenamefont
  {Blanchet}, \citenamefont {Faye}, \citenamefont {Seraille},\ and\
  \citenamefont {Trestini}}]{Blanchet:2026suq}%
  \BibitemOpen
  \bibfield  {author} {\bibinfo {author} {\bibfnamefont {L.}~\bibnamefont
  {Blanchet}}, \bibinfo {author} {\bibfnamefont {G.}~\bibnamefont {Faye}},
  \bibinfo {author} {\bibfnamefont {E.}~\bibnamefont {Seraille}},\ and\
  \bibinfo {author} {\bibfnamefont {D.}~\bibnamefont {Trestini}},\ }\href@noop
  {} {\  (\bibinfo {year} {2026})},\ \Eprint {https://arxiv.org/abs/2601.06743}
  {arXiv:2601.06743 [gr-qc]} \BibitemShut {NoStop}%
\bibitem [{\citenamefont {Driesse}\ \emph {et~al.}(2026)\citenamefont
  {Driesse}, \citenamefont {Jakobsen}, \citenamefont {Mogull}, \citenamefont
  {Nega}, \citenamefont {Plefka}, \citenamefont {Sauer},\ and\ \citenamefont
  {Usovitsch}}]{Driesse:2026qiz}%
  \BibitemOpen
  \bibfield  {author} {\bibinfo {author} {\bibfnamefont {M.}~\bibnamefont
  {Driesse}}, \bibinfo {author} {\bibfnamefont {G.~U.}\ \bibnamefont
  {Jakobsen}}, \bibinfo {author} {\bibfnamefont {G.}~\bibnamefont {Mogull}},
  \bibinfo {author} {\bibfnamefont {C.}~\bibnamefont {Nega}}, \bibinfo {author}
  {\bibfnamefont {J.}~\bibnamefont {Plefka}}, \bibinfo {author} {\bibfnamefont
  {B.}~\bibnamefont {Sauer}},\ and\ \bibinfo {author} {\bibfnamefont
  {J.}~\bibnamefont {Usovitsch}},\ }\href@noop {} {\  (\bibinfo {year}
  {2026})},\ \Eprint {https://arxiv.org/abs/2601.16256} {arXiv:2601.16256
  [hep-th]} \BibitemShut {NoStop}%
\bibitem [{\citenamefont {Goldberger}\ and\ \citenamefont
  {Rothstein}(2006{\natexlab{a}})}]{nrgr}%
  \BibitemOpen
  \bibfield  {author} {\bibinfo {author} {\bibfnamefont {W.~D.}\ \bibnamefont
  {Goldberger}}\ and\ \bibinfo {author} {\bibfnamefont {I.~Z.}\ \bibnamefont
  {Rothstein}},\ }\href {https://doi.org/10.1103/PhysRevD.73.104029} {\bibfield
   {journal} {\bibinfo  {journal} {Phys. Rev.}\ }\textbf {\bibinfo {volume}
  {D73}},\ \bibinfo {pages} {104029} (\bibinfo {year} {2006}{\natexlab{a}})},\
  \Eprint {https://arxiv.org/abs/hep-th/0409156} {arXiv:hep-th/0409156}
  \BibitemShut {NoStop}%
\bibitem [{\citenamefont {Porto}(2006)}]{nrgrs}%
  \BibitemOpen
  \bibfield  {author} {\bibinfo {author} {\bibfnamefont {R.~A.}\ \bibnamefont
  {Porto}},\ }\href {https://doi.org/10.1103/PhysRevD.73.104031} {\bibfield
  {journal} {\bibinfo  {journal} {Phys. Rev. D}\ }\textbf {\bibinfo {volume}
  {73}},\ \bibinfo {pages} {104031} (\bibinfo {year} {2006})},\ \Eprint
  {https://arxiv.org/abs/gr-qc/0511061} {arXiv:gr-qc/0511061} \BibitemShut
  {NoStop}%
\bibitem [{\citenamefont {Goldberger}\ and\ \citenamefont
  {Rothstein}(2006{\natexlab{b}})}]{dis1}%
  \BibitemOpen
  \bibfield  {author} {\bibinfo {author} {\bibfnamefont {W.}~\bibnamefont
  {Goldberger}}\ and\ \bibinfo {author} {\bibfnamefont {I.}~\bibnamefont
  {Rothstein}},\ }\href {https://doi.org/10.1103/PhysRevD.73.104030} {\bibfield
   {journal} {\bibinfo  {journal} {Phys. Rev. D}\ }\textbf {\bibinfo {volume}
  {73}},\ \bibinfo {pages} {104030} (\bibinfo {year} {2006}{\natexlab{b}})},\
  \Eprint {https://arxiv.org/abs/hep-th/0511133} {arXiv:hep-th/0511133}
  \BibitemShut {NoStop}%
\bibitem [{\citenamefont {Porto}(2008)}]{dis2}%
  \BibitemOpen
  \bibfield  {author} {\bibinfo {author} {\bibfnamefont {R.~A.}\ \bibnamefont
  {Porto}},\ }\href {https://doi.org/10.1103/PhysRevD.77.064026} {\bibfield
  {journal} {\bibinfo  {journal} {Phys. Rev. D}\ }\textbf {\bibinfo {volume}
  {77}},\ \bibinfo {pages} {064026} (\bibinfo {year} {2008})},\ \Eprint
  {https://arxiv.org/abs/0710.5150} {arXiv:0710.5150} \BibitemShut {NoStop}%
\bibitem [{\citenamefont {Porto}\ and\ \citenamefont
  {Rothstein}(2008{\natexlab{a}})}]{nrgrss}%
  \BibitemOpen
  \bibfield  {author} {\bibinfo {author} {\bibfnamefont {R.~A.}\ \bibnamefont
  {Porto}}\ and\ \bibinfo {author} {\bibfnamefont {I.~Z.}\ \bibnamefont
  {Rothstein}},\ }\href {https://doi.org/10.1103/PhysRevD.78.044012} {\bibfield
   {journal} {\bibinfo  {journal} {Phys.Rev.}\ }\textbf {\bibinfo {volume}
  {D78}},\ \bibinfo {pages} {044012} (\bibinfo {year} {2008}{\natexlab{a}})},\
  \Eprint {https://arxiv.org/abs/0802.0720} {arXiv:0802.0720} \BibitemShut
  {NoStop}%
\bibitem [{\citenamefont {Porto}\ and\ \citenamefont
  {Rothstein}(2008{\natexlab{b}})}]{nrgrs2}%
  \BibitemOpen
  \bibfield  {author} {\bibinfo {author} {\bibfnamefont {R.~A.}\ \bibnamefont
  {Porto}}\ and\ \bibinfo {author} {\bibfnamefont {I.~Z.}\ \bibnamefont
  {Rothstein}},\ }\href {https://doi.org/10.1103/PhysRevD.78.044013} {\bibfield
   {journal} {\bibinfo  {journal} {Phys.Rev.}\ }\textbf {\bibinfo {volume}
  {D78}},\ \bibinfo {pages} {044013} (\bibinfo {year} {2008}{\natexlab{b}})},\
  \Eprint {https://arxiv.org/abs/0804.0260} {arXiv:0804.0260} \BibitemShut
  {NoStop}%
\bibitem [{\citenamefont {Rothstein}(2014)}]{iragrg}%
  \BibitemOpen
  \bibfield  {author} {\bibinfo {author} {\bibfnamefont {I.~Z.}\ \bibnamefont
  {Rothstein}},\ }\href {https://doi.org/10.1007/s10714-014-1726-y} {\bibfield
  {journal} {\bibinfo  {journal} {Gen. Rel. Grav.}\ }\textbf {\bibinfo {volume}
  {46}},\ \bibinfo {pages} {1726} (\bibinfo {year} {2014})}\BibitemShut
  {NoStop}%
\bibitem [{\citenamefont {Porto}(2016)}]{review}%
  \BibitemOpen
  \bibfield  {author} {\bibinfo {author} {\bibfnamefont {R.~A.}\ \bibnamefont
  {Porto}},\ }\href {https://doi.org/10.1016/j.physrep.2016.04.003} {\bibfield
  {journal} {\bibinfo  {journal} {Phys. Rept.}\ }\textbf {\bibinfo {volume}
  {633}},\ \bibinfo {pages} {1} (\bibinfo {year} {2016})},\ \Eprint
  {https://arxiv.org/abs/1601.04914} {arXiv:1601.04914} \BibitemShut {NoStop}%
\bibitem [{\citenamefont {Goldberger}(2022)}]{Goldberger:2022ebt}%
  \BibitemOpen
  \bibfield  {author} {\bibinfo {author} {\bibfnamefont {W.~D.}\ \bibnamefont
  {Goldberger}},\ }\href@noop {} {\  (\bibinfo {year} {2022})},\ \Eprint
  {https://arxiv.org/abs/2206.14249} {arXiv:2206.14249 [hep-th]} \BibitemShut
  {NoStop}%
\bibitem [{\citenamefont {Porto}\ and\ \citenamefont {Riva}(2026)}]{memory2}%
  \BibitemOpen
  \bibfield  {author} {\bibinfo {author} {\bibfnamefont {R.~A.}\ \bibnamefont
  {Porto}}\ and\ \bibinfo {author} {\bibfnamefont {M.~M.}\ \bibnamefont
  {Riva}},\ }\href@noop {} {\  (\bibinfo {year} {2026})},\ \Eprint
  {https://arxiv.org/abs/2604.09545} {arXiv:2604.09545 [gr-qc]} \BibitemShut
  {NoStop}%
\bibitem [{\citenamefont {Brunello}\ \emph {et~al.}(2025)\citenamefont
  {Brunello}, \citenamefont {Mandal}, \citenamefont {Mastrolia}, \citenamefont
  {Patil}, \citenamefont {Pegorin}, \citenamefont {Ronca}, \citenamefont
  {Smith}, \citenamefont {Steinhoff},\ and\ \citenamefont
  {Torres~Bobadilla}}]{Brunello:2025gpf}%
  \BibitemOpen
  \bibfield  {author} {\bibinfo {author} {\bibfnamefont {G.}~\bibnamefont
  {Brunello}}, \bibinfo {author} {\bibfnamefont {M.~K.}\ \bibnamefont
  {Mandal}}, \bibinfo {author} {\bibfnamefont {P.}~\bibnamefont {Mastrolia}},
  \bibinfo {author} {\bibfnamefont {R.}~\bibnamefont {Patil}}, \bibinfo
  {author} {\bibfnamefont {M.}~\bibnamefont {Pegorin}}, \bibinfo {author}
  {\bibfnamefont {J.}~\bibnamefont {Ronca}}, \bibinfo {author} {\bibfnamefont
  {S.}~\bibnamefont {Smith}}, \bibinfo {author} {\bibfnamefont
  {J.}~\bibnamefont {Steinhoff}},\ and\ \bibinfo {author} {\bibfnamefont
  {W.~J.}\ \bibnamefont {Torres~Bobadilla}},\ }\href@noop {} {\  (\bibinfo
  {year} {2025})},\ \Eprint {https://arxiv.org/abs/2512.19498}
  {arXiv:2512.19498 [hep-th]} \BibitemShut {NoStop}%
\bibitem [{\citenamefont {Jakobsen}\ \emph
  {et~al.}(2023{\natexlab{a}})\citenamefont {Jakobsen}, \citenamefont {Mogull},
  \citenamefont {Plefka},\ and\ \citenamefont {Sauer}}]{Jakobsen:2023hig}%
  \BibitemOpen
  \bibfield  {author} {\bibinfo {author} {\bibfnamefont {G.~U.}\ \bibnamefont
  {Jakobsen}}, \bibinfo {author} {\bibfnamefont {G.}~\bibnamefont {Mogull}},
  \bibinfo {author} {\bibfnamefont {J.}~\bibnamefont {Plefka}},\ and\ \bibinfo
  {author} {\bibfnamefont {B.}~\bibnamefont {Sauer}},\ }\href
  {https://doi.org/10.1103/PhysRevLett.131.241402} {\bibfield  {journal}
  {\bibinfo  {journal} {Phys. Rev. Lett.}\ }\textbf {\bibinfo {volume} {131}},\
  \bibinfo {pages} {241402} (\bibinfo {year} {2023}{\natexlab{a}})},\ \Eprint
  {https://arxiv.org/abs/2308.11514} {arXiv:2308.11514 [hep-th]} \BibitemShut
  {NoStop}%
\bibitem [{\citenamefont {Jakobsen}\ \emph
  {et~al.}(2023{\natexlab{b}})\citenamefont {Jakobsen}, \citenamefont {Mogull},
  \citenamefont {Plefka}, \citenamefont {Sauer},\ and\ \citenamefont
  {Xu}}]{Jakobsen:2023ndj}%
  \BibitemOpen
  \bibfield  {author} {\bibinfo {author} {\bibfnamefont {G.~U.}\ \bibnamefont
  {Jakobsen}}, \bibinfo {author} {\bibfnamefont {G.}~\bibnamefont {Mogull}},
  \bibinfo {author} {\bibfnamefont {J.}~\bibnamefont {Plefka}}, \bibinfo
  {author} {\bibfnamefont {B.}~\bibnamefont {Sauer}},\ and\ \bibinfo {author}
  {\bibfnamefont {Y.}~\bibnamefont {Xu}},\ }\href
  {https://doi.org/10.1103/PhysRevLett.131.151401} {\bibfield  {journal}
  {\bibinfo  {journal} {Phys. Rev. Lett.}\ }\textbf {\bibinfo {volume} {131}},\
  \bibinfo {pages} {151401} (\bibinfo {year} {2023}{\natexlab{b}})},\ \Eprint
  {https://arxiv.org/abs/2306.01714} {arXiv:2306.01714 [hep-th]} \BibitemShut
  {NoStop}%
\bibitem [{\citenamefont {Lange}\ \emph {et~al.}(2026)\citenamefont {Lange},
  \citenamefont {Usovitsch},\ and\ \citenamefont {Wu}}]{Lange:2025fba}%
  \BibitemOpen
  \bibfield  {author} {\bibinfo {author} {\bibfnamefont {F.}~\bibnamefont
  {Lange}}, \bibinfo {author} {\bibfnamefont {J.}~\bibnamefont {Usovitsch}},\
  and\ \bibinfo {author} {\bibfnamefont {Z.}~\bibnamefont {Wu}},\ }\href
  {https://doi.org/10.1016/j.cpc.2025.109999} {\bibfield  {journal} {\bibinfo
  {journal} {Comput. Phys. Commun.}\ }\textbf {\bibinfo {volume} {322}},\
  \bibinfo {pages} {109999} (\bibinfo {year} {2026})},\ \Eprint
  {https://arxiv.org/abs/2505.20197} {arXiv:2505.20197 [hep-ph]} \BibitemShut
  {NoStop}%
\bibitem [{\citenamefont {Smirnov}\ and\ \citenamefont
  {Zeng}(2025)}]{Smirnov:2025prc}%
  \BibitemOpen
  \bibfield  {author} {\bibinfo {author} {\bibfnamefont {A.~V.}\ \bibnamefont
  {Smirnov}}\ and\ \bibinfo {author} {\bibfnamefont {M.}~\bibnamefont {Zeng}},\
  }\href@noop {} {\  (\bibinfo {year} {2025})},\ \Eprint
  {https://arxiv.org/abs/2510.07150} {arXiv:2510.07150 [hep-ph]} \BibitemShut
  {NoStop}%
\bibitem [{\citenamefont {Lee}(2014)}]{Lee:2013mka}%
  \BibitemOpen
  \bibfield  {author} {\bibinfo {author} {\bibfnamefont {R.~N.}\ \bibnamefont
  {Lee}},\ }\href {https://doi.org/10.1088/1742-6596/523/1/012059} {\bibfield
  {journal} {\bibinfo  {journal} {J. Phys. Conf. Ser.}\ }\textbf {\bibinfo
  {volume} {523}},\ \bibinfo {pages} {012059} (\bibinfo {year} {2014})},\
  \Eprint {https://arxiv.org/abs/1310.1145} {arXiv:1310.1145 [hep-ph]}
  \BibitemShut {NoStop}%
\bibitem [{\citenamefont {de~la Cruz}\ and\ \citenamefont
  {Kosower}(2026)}]{delaCruz:2026mas}%
  \BibitemOpen
  \bibfield  {author} {\bibinfo {author} {\bibfnamefont {L.}~\bibnamefont
  {de~la Cruz}}\ and\ \bibinfo {author} {\bibfnamefont {D.~A.}\ \bibnamefont
  {Kosower}},\ }\href@noop {} {\  (\bibinfo {year} {2026})},\ \Eprint
  {https://arxiv.org/abs/2602.22111} {arXiv:2602.22111 [hep-ph]} \BibitemShut
  {NoStop}%
\bibitem [{\citenamefont {Henn}(2013)}]{Henn:2013pwa}%
  \BibitemOpen
  \bibfield  {author} {\bibinfo {author} {\bibfnamefont {J.~M.}\ \bibnamefont
  {Henn}},\ }\href {https://doi.org/10.1103/PhysRevLett.110.251601} {\bibfield
  {journal} {\bibinfo  {journal} {Phys. Rev. Lett.}\ }\textbf {\bibinfo
  {volume} {110}},\ \bibinfo {pages} {251601} (\bibinfo {year} {2013})},\
  \Eprint {https://arxiv.org/abs/1304.1806} {arXiv:1304.1806} \BibitemShut
  {NoStop}%
\bibitem [{\citenamefont {Jinno}\ \emph {et~al.}(2022)\citenamefont {Jinno},
  \citenamefont {K\"alin}, \citenamefont {Liu},\ and\ \citenamefont
  {Rubira}}]{Jinno:2022sbr}%
  \BibitemOpen
  \bibfield  {author} {\bibinfo {author} {\bibfnamefont {R.}~\bibnamefont
  {Jinno}}, \bibinfo {author} {\bibfnamefont {G.}~\bibnamefont {K\"alin}},
  \bibinfo {author} {\bibfnamefont {Z.}~\bibnamefont {Liu}},\ and\ \bibinfo
  {author} {\bibfnamefont {H.}~\bibnamefont {Rubira}},\ }\href@noop {} {\
  (\bibinfo {year} {2022})},\ \Eprint {https://arxiv.org/abs/2209.01091}
  {arXiv:2209.01091 [hep-th]} \BibitemShut {NoStop}%
\bibitem [{\citenamefont {Bailey}\ and\ \citenamefont
  {Ferguson}(1991)}]{bailey1991polynomial}%
  \BibitemOpen
  \bibfield  {author} {\bibinfo {author} {\bibfnamefont {D.}~\bibnamefont
  {Bailey}}\ and\ \bibinfo {author} {\bibfnamefont {H.}~\bibnamefont
  {Ferguson}},\ }\href@noop {} {\bibfield  {journal} {\bibinfo  {journal} {NASA
  Technical Report RNR-91-032}\ } (\bibinfo {year} {1991})}\BibitemShut
  {NoStop}%
\bibitem [{\citenamefont {Bailey}\ and\ \citenamefont
  {Broadhurst}(2001)}]{Bailey:1999nv}%
  \BibitemOpen
  \bibfield  {author} {\bibinfo {author} {\bibfnamefont {D.~H.}\ \bibnamefont
  {Bailey}}\ and\ \bibinfo {author} {\bibfnamefont {D.~J.}\ \bibnamefont
  {Broadhurst}},\ }\href {https://doi.org/10.1090/S0025-5718-00-01278-3}
  {\bibfield  {journal} {\bibinfo  {journal} {Math. Comput.}\ }\textbf
  {\bibinfo {volume} {70}},\ \bibinfo {pages} {1719} (\bibinfo {year}
  {2001})},\ \Eprint {https://arxiv.org/abs/math/9905048} {arXiv:math/9905048}
  \BibitemShut {NoStop}%
\end{thebibliography}%

\end{document}